\documentclass[acmsmall,screen]{acmart}

\usepackage{booktabs}   
\usepackage{subcaption} 

\usepackage[utf8]{inputenc}
\usepackage[T1]{fontenc}

\usepackage{amsmath, amssymb}
\usepackage{array}
\usepackage{comment}
\usepackage{enumitem}
\usepackage{graphicx}
\usepackage{listings}
\usepackage{microtype}
\usepackage{multirow}
\usepackage{xcolor, colortbl}
\usepackage{wrapfig}
\usepackage{soul}

\newcommand{\toolname}{\textsc{Plan\-Alyzer}}
\newcommand{\planout}{\textsc{Plan\-Out}}

\newcommand{\companyname}{Facebook}

\definecolor{mygreen}{rgb}{0,0.6,0}
\definecolor{mygray}{rgb}{0.5,0.5,0.5}
\definecolor{mymauve}{rgb}{0.58,0,0.82}
\definecolor{myred}{rgb}{0.6,0.1,0.1}
\newcommand{\incolcode}{\ttfamily\scriptsize}

\newcommand{\numuniqueexps}[0]{240}

\author{Emma Tosch}
\email{etosch@cs.umass.edu}
\affiliation{
  \institution{CICS, University of Massachusetts Amherst}
  \streetaddress{140 Governors Drive}
  \city{Amherst}
  \state{MA}
  \postcode{01002}
  \country{USA}
}
\author{Eytan Bakshy}
\email{ebakshy@fb.com}
\affiliation{
  \institution{Facebook, Inc.}
  \streetaddress{1 Hacker Way}
  \city{Menlo Park}
  \state{CA}
  \postcode{94025}
  \country{USA}
}
\author{Emery D. Berger}
\email{emery@cs.umass.edu}
\author{David D. Jensen}
\email{jensen@cs.umass.edu}
\author{J. Eliot B. Moss}
\email{moss@cs.umass.edu}
\affiliation{
  \institution{CICS, University of Massachusetts Amherst}
  \country{USA}
}

\setcopyright{rightsretained}
\acmPrice{}
\acmDOI{10.1145/3360608}
\acmYear{2019}
\copyrightyear{2019}
\acmJournal{PACMPL}
\acmVolume{3}
\acmNumber{OOPSLA}
\acmArticle{182}
\acmMonth{10}

\begin{document}
\title{\toolname{}:  Assessing Threats to the Validity of Online Experiments}

\begin{abstract}

Online experiments have become a ubiquitous aspect of design and engineering processes within Internet firms.  
As the scale of experiments has grown, so has the complexity of their design and implementation.  In response, firms have developed software frameworks for designing and deploying online experiments.  Ensuring that experiments in these frameworks are correctly designed and that their results are trustworthy---referred to as \emph{internal validity}---can be difficult.  Currently, verifying internal validity requires manual inspection by someone with substantial expertise in experimental design.

We present the first approach for statically checking the internal validity of online experiments. Our checks are based on well-known problems that arise in experimental design and causal inference. Our analyses target \planout{}, a widely deployed, open-source experimentation framework that uses a domain-specific language to specify and run complex experiments.  We have built a tool called \toolname{} that checks \planout{} programs for a variety of threats to internal validity, including failures of randomization, treatment assignment, and causal sufficiency. \toolname{} uses its analyses to automatically generate \emph{contrasts}, a key type of information required to perform valid statistical analyses over the results of these experiments. We demonstrate \toolname{}'s utility on a corpus of \planout{} scripts deployed in production at \companyname{}, and we evaluate its ability to identify threats to validity on a mutated subset of this corpus. \toolname{} has both precision and recall of 92\% on the mutated corpus, and 82\% of the contrasts it generates match hand-specified data.

\end{abstract}

\ccsdesc[500]{Software and its engineering~Scripting languages}
\ccsdesc[500]{Software and its engineering~Domain specific languages}
\ccsdesc[500]{Software and its engineering~Software testing and debugging}
\ccsdesc[300]{Software and its engineering~Specialized application languages}
\ccsdesc[300]{Software and its engineering~Empirical software validation}

\keywords{Experimental Design, Threats to Validity, Online Experiments}

\maketitle


\section{Introduction}
\label{sec:introduction}
Many organizations conduct online experiments to assist decision-making, and many of the largest organizations now employ some form of federated experimentation management ~\cite{Bakshy:2014:DDO:2566486.2567967, tang2010overlapping,
kohavi2009controlled, crook2009seven, tang2015holistic}. These systems often include software components that make designing experiments easier or that automatically monitor experimental results. 
One popular example component of such experimentation systems is Facebook's \planout{}: a domain-specific language for experimental design.\footnote{\url{http://facebook.github.io/planout}}  

The state of the art for validating \emph{experimental designs} (i.e., the procedure for conducting an experiment) is manual human review. The most common experimental design on the Web is the A/B test, where users see one of two variants (A and B). The state of the art for analyzing the \emph{results} of experimental designs depends on the design: for A/B tests, the outcomes of interest (e.g., click rates) can be computed and compared automatically, but more sophisticated designs require specialized analysis. Many  experiments written in a DSL such as \planout{} can be cumbersome to validate, and they cannot be analyzed using existing automated methods.

\paragraph{Validation of Experimental Designs.} Experiments expressed as programs can have errors that are unique to the intersection of experimentation and software. 
\citeN{shadish002experimental} enumerate a taxonomy of nine well-understood design errors in the experimental design literature, referred to as \emph{threats to internal validity}---i.e., the degree to which valid causal conclusions can be drawn within the context of the study. Seven of these errors can be avoided when the researcher employs a \emph{randomized experiment} that behaves as expected.\footnote{The two remaining threats to validity that are \emph{not} obviated by randomization are \emph{attrition}, described in \S\ref{sec:threats}, and \emph{testing}. Testing in experimental design refers to taking an initial measurement and then using the test instrument to conduct an experiment. Analysis may not be able to differentiate between the effect that a test was designed to measure and the effect of subjects learning the test itself. Testing is a form of \emph{within-subjects} analysis that is not typically employed in online field experiments and whose analyses are outside the scope of this work.} Therefore, ensuring randomization is critical to internal validity. However, randomization failures in programs manifest differently from randomization failures in the physical world: for example, a program cannot disobey an experimental protocol, but data flow can break randomization if a probability is erroneously set to zero.

\paragraph{Validation of Statistical Conclusions.} Control-flow operators, calls to external services, and in-language mechanisms for data recording prohibit simple automatic variable monitoring. For example, an experiment that sets variables differently on the basis of the current country of the user cannot na{\"i}vely aggregate results across all participants in the experiment. Such an experiment would require additional adjustment during post-experiment analysis to account for the potential confounding introduced by a user's current country.


\paragraph{Contributions.} The work presented in this paper: (1) statically identifies sources of statistical bias in programmatically defined experiments, and (2) presents methods for automatically generating 
the most common statistical analysis 
for the largest class of such experiments. 
We make the following contributions:
\begin{enumerate}[label=\textbf{C\arabic*}, leftmargin=*]
\item\label{contrib:tool} \textbf{Software for the static analysis of  experiments.} We introduce \toolname{}, the first tool, to our knowledge, for analyzing online experiments statically~(\S\ref{sec:tool}).\footnote{\url{http://www.github.com/KDL-UMass/PlanAlyzer}} 
\toolname{} produces three key pieces of information: (1) a list of the variables in the environment that are actually being randomly assigned; (2) the variables that are recorded for analysis; and (3) the variables that may be legitimately compared when computing causal effects.
These three pieces of information are required in order to determine whether there are any valid statistical analyses of the recorded results of an experiment, and, when possible, what those analyses are.

\item\label{contrib:errors} \textbf{Characterizing errors and bad practices (code smells) unique to programmatically defined experiments.} Traditional errors in offline experimentation can take on unusual forms in programmatically defined experiments. Additionally, some coding practices can lead to faults during downstream statistical analysis, highlighting the potential utility of defining code smells for experiments~\cite{fowler2018refactoring}.  In \S\ref{sec:bugs}, we introduce errors and code smells
that
arise from the intersection of experiments and software. 
\item\label{contrib:analysis} \textbf{Empirical analysis of real experiments.} We report \toolname{}'s performance on a corpus of real-world \planout{} scripts provided by \companyname{} (\S\ref{sec:corpus}-\ref{sec:evaluation}). 
Due to the vetting process at \companyname{}, few errors exist naturally in the corpus. Therefore, we perform mutation analysis to approximate a real-world distribution of errors.
\companyname{} also provided a corpus of human-generated \emph{contrasts} (the set of variable values that that are allowed to be compared, necessary for estimating causal effect).\footnote{Appendix~\ref{app:glossary} provides a glossary for the terminology of experimental design.} We demonstrate \toolname{}'s effectiveness in finding major threats to validity and in automatically generating contrasts.
\end{enumerate}

\paragraph{Organization.} We begin with a motivating example experiment in Section~\ref{sec:example}.
Section~\ref{sec:background} gives background in experimental design and causal inference (\S\ref{sec:terminology}), outlines common threats to validity in causal inference (\S\ref{sec:threats}), and presents the \planout{} framework (\S\ref{sec:online-experimentation}). Section~\ref{sec:tool}--Section~\ref{sec:corpus} describe our contributions and Section~~\ref{sec:evaluation} presents our evaluation. We discuss related work in Section~\ref{sec:related} and summarize our findings in Section~\ref{sec:conclusions}. We also include a glossary of relevant terminology in Appendix~\ref{app:glossary} and a discussion of features that make analyses in \planout{} nontrivial in Appendix~\ref{app:improvements}.

\section{Motivating Example}
\label{sec:example}
Consider an engineering team aiming improve the quality of video streaming for users of its mobile streaming service and thus increase its usage~\cite{krishnan2013video}. Users access the service over heterogeneous networks, and one way to improve quality is to change the video bit rate; the firm cannot control network bandwidth, but they can control how much data to transmit per second. Streaming video at higher bit rates will result in higher quality video, at the expense of increased data use. The team knows it wants to incorporate location information into how they determine bandwidth, but they disagree about what location information to use, and how to implement their design. They discuss two alternatives: (1) assign users to bit rates based on country, which has previously proven to be effective; or (2) assign users to bit rates based on real-time estimates of network latency, which has never been tested.

\paragraph{Country-Level Design.} The team considers two different bit rates: 400 kbit/s (low) and 750 kbit/s (high). Streaming videos at high bit rates may result in poor performance for individuals in emerging markets, so the team chooses to allocate fewer users to the high bit rate within these markets. Furthermore, the team chooses to constrain the population to markets the team understands well, and for which they have a large quantity of data (e.g., India, Brazil, the US, and Canada). 

\paragraph{Dynamic Design.} Users may change networks throughout the day; each of these networks may have different levels of latency. Therefore, the team considers a design that uses a personalized dynamic treatment regime, which maps users to different bit rates depending on network connection type~\cite{murphy2003optimal}. 

\vspace{1em}

The country-level design is similar to a classic A/B test over the possible max bit rates, per country: it is straightforward and should return results quickly. However, it is much more constrained than the dynamic design, using country as a coarse-grained proxy for bandwidth availability. The dynamic design takes into account contextual information that may change over the course of the experiment, causing the devices to receive different treatments at different times. Given the tradeoffs between these two experiments, the team decides to randomly assign participants to either one experiment or the other.

\begin{figure}
\centering
\begin{minipage}{\linewidth}
\begin{lstlisting}[basicstyle=\incolcode, numberstyle=\tiny, numbers=left, xleftmargin=2.5em]
dynamic_policy = bernoulliTrial(p=0.3, unit=userid);
context = getContext(deviceid=deviceid, userid=userid);
country = getUserCountry(userid=userid);
emerging_market = (country == 'IN') || (country == 'BR');
established_market = (country == 'US') || (country == 'CA');
if(dynamic_policy) {
    weights = getBanditWeights(context=context);
    choices = getBanditChoices(context=context);
    max_bitrate = weightedChoice(choices=choices, weights=weights, unit=userid);
} else {
    if (emerging_market) {
      max_bitrate = weightedChoice(choices=[400, 750], weights=[0.9, 0.1], unit=userid);
    } else if (established_market) {
      max_bitrate = weightedChoice(choices=[400, 750], weights=[0.5, 0.5], unit=userid);
    } else {
      max_bitrate = 400;
    }
}
\end{lstlisting}
\end{minipage}
\caption{\label{code:example-bitrate-planout} The example experiment of \S\ref{sec:example}, written in the \planout{} language. This experiment aims to improve video streaming experiences for users with potentially heterogeneous network conditions by testing different bit rates for video transmission in a mobile application software, according to a randomly chosen policy. One policy is dynamic; treatments depend on user features, potentially varying over the course of the experiment. As discussed in \S\ref{sec:online-experimentation}, \planout{} programs are executed within a containing environment. 
\lstinline{getContext}, \lstinline{getUserCountry}, \lstinline{getBanditWeights}, and \lstinline{getBanditChoices} are all functions that exist within that environment. }
\end{figure}

\begin{figure}
  \begin{scriptsize}
    \begin{verbatim}Conditioning set: {}
============
Avg(Y|dynamic_policy=true) - Avg(Y|dynamic_policy=false)

Conditioning set:
{emerging_market : true} 
============
Avg(Y|max_bitrate=400, dynamic_policy=false) - Avg(Y|max_bitrate=750, dynamic_policy=false) 

Conditioning set:
{emerging_market : false; established_market : true}
============
Avg(Y|max_bitrate=400, dynamic_policy=false) - Avg(Y|max_bitrate=750, dynamic_policy=false)\end{verbatim}
\end{scriptsize}
\caption{\label{fig:example-bitrate-analysis}Valid contrasts and their associated conditioning sets (i.e., constraints on the analyses), presented using one output option for \toolname{} (\ref{contrib:tool}), for the program in Figure~\ref{code:example-bitrate-planout}. The outcome of interest is the average percentage of videos watched. 
As we discuss in \S\ref{sec:background}, we generally abstract over the outcome and simply refer to it as $Y$. 
\toolname{} finds three conditioning sets, each having one set of valid contrasts. 
A naive approach to comparing bit rates might average over all data points. If this were valid, \toolname{} would have produced \texttt{Avg(Y|max\_bitrate=400)-Avg(Y|max\_bitrate=750)} for the empty conditioning set. 
}
\end{figure}

The script in Figure~\ref{code:example-bitrate-planout} depicts one way of representing these experiments in \planout{}.  There are four paths through the program, and three of them randomly assign values to \texttt{max\_bitrate} directly. The fourth path can only be said to randomly assign \texttt{max\_bitrate} indirectly, via random branching.

Figure~\ref{fig:example-bitrate-analysis} depicts \toolname{}'s output in a human-readable format. This output specifies the \emph{valid contrasts}, or the variables that may be legitimately compared. While there is a relationship between paths through a program and contrasts, the mapping is imperfect: only the latter two of the three contrasts correspond to a path. 

Each contrast has two components: a possibly empty \emph{conditioning set}, which corresponds to a kind of constraint on analysis, and a list of valid pairwise comparisons: only cases for which
   the constraint holds should be used when analyzing the associated
   contrast(s).\footnote{Note that our default output format is a collection of comma-separated value files (CSVs) that partition the contrasts by those that may be compared. We expect end-users to load these CSVs into a database. Producing the full list of pairwise comparisons when there are a large number of variables to compare is inadvisable. We produce the format presented here because it is the standard way of expressing contrasts in the experimental design community.} The first contrast corresponds to comparing between the two approaches. This type of comparison may not actually be of interest to the team, but it is valid. The second and third contrasts must be analyzed separately: whether a user is in an emerging market or an established market determines their probability of being assigned the high or low bit rate, but the market may also have an influence on the average percentage of videos watched ($Y$). Therefore, na{\"i}vely aggregating over the high and low bit rates for the country-based experiment would not be correct.

Now consider an alternative version of this program.  Assume that we have the functions \texttt{in\-Emerging\-Market} and \texttt{in\-Established\-Market} available and that these functions take no arguments, but use the current location of the device to determine whether the user is currently in the appropriate market.  Then we might replace the call to \texttt{getUserCountry} and lines 11-17 with:

\begin{lstlisting}[basicstyle=\incolcode]
if (inEmergingMarket()) {
    max_bitrate = weightedChoice(choices=[400, 750], weights=[0.9, 0.1], unit=userid);
} else if (inEstablishedMarket()) {
    max_bitrate = weightedChoice(choices=[400, 750], weights=[0.5, 0.5], unit=userid);
} else {
    max_bitrate = 400;
}
\end{lstlisting}

The results of the calls to \texttt{inEmergingMarket} and \texttt{inEstablishedMarket} would then be stored in intermediate values and therefore would not be recorded. The logic of experimentation remains the same, but now we have no way to recover which market users were in from the recorded data. In this case, the only valid contrast would be the first one listed in Figure~\ref{fig:example-bitrate-analysis}. If this were a standalone experiment, random assignment would be completely broken.

\section{Background}
\label{sec:background}
Experiments are procedures designed to measure the effects of interventions. Two common elements of experimentation are \emph{randomization}, in which experimental \emph{units} (e.g., users or devices) are randomly assigned to \emph{treatments}, and \emph{contrasts}, in which some set of outcomes are compared between units in each treatment group. One part of the example experiment shown in Figure~\ref{code:example-bitrate-planout} randomly assigns users to different bit rates (\textit{treatments}).  The other part of the experiment assigns users to either a dynamic or a country-based policy. The outcome might be the amount of time spent watching videos.

\subsection{Experimental Design and Causal Inference}
\label{sec:terminology}
Experimental design refers to the different methods of deliberately assigning treatments to units, in service of estimating the effect of those treatments on an outcome of interest that is typically denoted as $Y$. In the example of Fig~\ref{code:example-bitrate-planout}, $Y$ corresponds to the average amount of time people spend watching videos. One set of treatments ($T$) is \lstinline{dynamic_policy=false}, which we will encode as 0, and \lstinline{dynamic_policy=true}, which we will encode as 1. Then $Y^{T=1}$ corresponds to the outcome for users receiving dynamic treatment assignment. 

The scope of this work applies to \emph{field experiments}, named for their origins in agriculture, where weather and ecosystems provide myriad additional variables that can influence experimental outcomes~\cite{fisher1936design}. 
 Ascertaining effects from a blend of human behavior and computerized systems requires researchers to treat these experiments as field experiments, rather than laboratory experiments, which are more common in computer science research~\cite{Blackburn:2008:WUS:1378704.1378723}. 
 Section~\ref{sec:online-experimentation} describes some of the technical difficulties associated with running experiments on complex, heterogeneous systems.

The collection of experimental units is the \emph{sample} (e.g., users, cookies, or devices). For a sufficiently large sample, random assignment ensures that the effects of any other variables on outcome are equally distributed across treatment groups, allowing accurate estimation of causal effects.

In addition to recording outcome data, an online experimentation system must also record the treatment actually assigned, rather than just the intended treatment. Practitioners often refer to this data collection process as \emph{logging}, but we will refer to it as \emph{recording}, to avoid overloading existing terminology in the software engineering literature. 

The function that estimates causal effect may take many forms. Nearly all such functions can be distilled into estimating the true difference between (1) an outcome under one treatment and (2) its \emph{potential outcome(s)} under another treatment. 

\paragraph{Average Treatment Effect (ATE)} 
In the case of a randomized experiment, we wish to know how $Y$ is affected by a treatment $T$. If $T$ is assigned completely at random, for example, according to:
\begin{center}
\begin{minipage}{0.6\textwidth}
\begin{lstlisting}[basicstyle=\incolcode]
T = uniformChoice(choices=[400, 750], unit=userid);
\end{lstlisting}
\end{minipage}
\end{center}
\noindent then the causal effect of $T$ (the \emph{average treatment effect}) can be estimated by simply taking the difference of the average outcome for units assigned to $T=1$ and $T=0$: $Avg(Y|T=1) - Avg(Y|T=0)$.  Such an experiment could be useful for learning how video watch time differs for equivalent individuals experiencing videos at the 400 or 750kbps setting.

It is not uncommon to use different probabilities for different kinds of users, such that some observed subgroup $S$ causes us to assign users to treatments with different probabilities.  This is illustrated in Figure~\ref{code:example-bitrate-planout}: a user's country, and therefore \lstinline{emerging_market}, will depend on some property of the population of users.  If we did not consider \lstinline{emerging_markets} in constructing our treatment control contrast, shown in Figure~\ref{fig:example-bitrate-analysis}, then our analysis would be incorrect. We can still estimate causal effects, but must instead compute the difference in means separately for different values of the variables in $S$. This is often referred to as \textit{subgroup analysis}. 

Were we to compare average percentage of videos watched for users outside India, Brazil, the US, or Canada with the average percentage of videos watched for users within these markets, our estimates of causal effect of \lstinline{max_bitrate} on abandonment would almost certainly be biased; the way lines 9-17 are written almost guarantees this. However, if we condition on the variables \lstinline{emerging_market} and \lstinline{established_market}, it becomes valid to estimate \lstinline{max_bitrate}. This estimand is known as the \emph{conditional average treatment effect} (CATE). In this paper, we will refer to ATE and CATE as both belonging to the same family of estimators, since ATE is a specific case of CATE (i.e., one with an empty conditioning set). Average effect estimators over finite sets of treatments can be expressed in terms of their valid contrasts.
\subsection{Threats to Validity of Causal Inference}
\label{sec:threats}
Three of the most serious threats to validity in causal inference are  \emph{selection bias}, \emph{confounding}, and a \emph{lack of positivity}~\cite{hernan2016causalinference}. Some instances of these threats can be readily identified in programmatically defined experiments.

\paragraph{Selection Bias.} Selection bias occurs when the sample chosen for treatment and analysis in the experiment differs in a systematic way from the underlying population.  This bias can arise due to a variety of reasons:
\begin{itemize}[leftmargin=*, itemsep=1em]
    \item \emph{Non-proportional sampling} causes selection bias by sampling from a subgroup that does not resemble the population. \citeN{wang2015forecasting} showed how to correct for this bias during \emph{post-hoc} analysis of political polling data from XBox users; the correction for a non-proportional sample requires the researcher to have access to a model of how the sampling population differs from the target population. When online experiments seek to answer questions about a platform's user base (e.g. hypotheses about user engagement with ads), non-proportional sampling is not likely to be an issue. 
    When researchers seek to extend hypotheses to the general population or offline behaviors, they may find that their sample differs in unexpected ways. For example, researchers across many disciplines use Amazon's Mechanical Turk (AMT) to collect data. \citeN{ahler2019micro} have shown that while both online and offline subjects engage in satisficing and trolling behavior, online respondents can spoof IP addresses, faking locations, while participating in studies more than once. These subjects pose a greater threat to the validity of AMT results than, e.g., random respondents or bots, due to their non-uniform (and therefore biased) behavior.
     
    \vspace{0.5em}
    Although non-proportional sampling is a cause of selection bias, in an online context it is tied to hypotheses, reasoning over which is currently out of scope of this work. We see an opportunity for future work in joint reasoning about hypotheses (e.g. predicates over the distribution of units, treatments, and the outcome variable $Y$) in order to truly validate programmatically defined experiments. Doing so requires additional information about the population, any proxy variables used in the hypotheses (e.g., using ``sad'' words denotes a sad mood), and ground-truth hypotheses from researchers. We know of no experimentation system that includes formally specified hypotheses.
    
    \item A failure causes complete loss of data between the assignment of units to treatment and the recording of results (known as \emph{dropout} or \emph{attrition}). Analogous behavior for online experiments may not be entirely relevant---e.g., were users to disable JavaScript or quit a social network, they may no longer be considered part of the population.  On the other hand, if, for example, during a storm a data center were to lose power and an experiment not be deployed to users within in particular region, we would have a form of dropout.
    
    \vspace{0.5em}
    
    Since dropout, by definition, only occurs after treatment assignment, recognizing it would require a form of dynamic analysis and thus it is outside the scope of this paper.

    \item During the mapping from subjects to treatments, randomization may fail. In offline contexts, this could be due to a breakdown in the experimental protocol. For example, a post-publication analysis of a large-scale randomized field experiment regarding the effects of the Mediterranean diet revealed that one of the investigators improperly assigned an entire village to the same treatment, despite an experimental protocol indicating that individuals be randomly assigned treatment~\cite{estruch2013primary, martinez2014dietary, mccook2018errors}. Although such deviations from the experimental protocol are unlikely to occur in programmatically defined experiments, it is always possible that there are flaws in the program itself that cause deterministic, and therefore biased, assignment of treatments to subjects. 

    \vspace{0.5em}
    This type of selection bias can be detected statically. In Section~\ref{sec:bugs}, we expand on biases of this variety, and describe how they may be expressed and detected in \planout{} programs.     

\end{itemize}

\paragraph{Confounding} Confounding occurs when there exists a variable that causes both treatment and outcome. For example, a common type of observational study of epidemiological effects compares between medicines prescribed, but the underlying disease causes both the choice of medicine and the subsequent symptoms~\cite{signorello2002confounding}. Confounding is a major threat to validity in observational studies, where researchers have no control over treatment assignment. 

Randomized treatment assignment ensures that there are no confounding variables.  However, a fully randomized experiment is not always possible or desirable.  In some cases, random assignment can be conditioned on some feature of the unit.  This can introduce confounding, especially when an experimentation script is sufficiently complex or where the feature that influenced assignment is not recorded. As long as the confounding variable can be measured, its effect may be factored out at analysis time, provided the analyst properly adjusts for it~\cite{Pearl:2009:CMR:1642718}. 


\paragraph{Lack of Positivity} When a treatment has some nonzero likelihood of being
observed in a data set for each combination of relevant variable values (i.e., across \emph{covariates}), it has positivity. A lack of positivity can lead to statistical bias, especially during subgroup analysis. Suppose you are conducting a drug trial for breast cancer. Due to suspected heterogeneity of the efficacy of the drug by sex, you analyze the data for men and women separately. However, your sample has no men, due to the significantly lower rate of breast cancer among men~\cite{ACSbreastcancerwomen, ACSbreastcancermen}. In this case, your study lacks positivity among men, since they have zero probability of treatment.

While an experimenter can sometimes correct for a lack of positivity in downstream statistical analysis, they must first be able to detect it.
Positivity is not typically considered an issue in experiments, since
experimenters are thought to completely control the assignment process.
However, in the context of online experiments, there are
some nontraditional and surprising reasons why positivity may not be achieved. Some of these reasons can be detected statically (\S\ref{sec:bugs}).

\subsection{The \planout{} Framework}
\label{sec:online-experimentation}
\begin{figure}
\centering
    \includegraphics[width=0.65\textwidth]{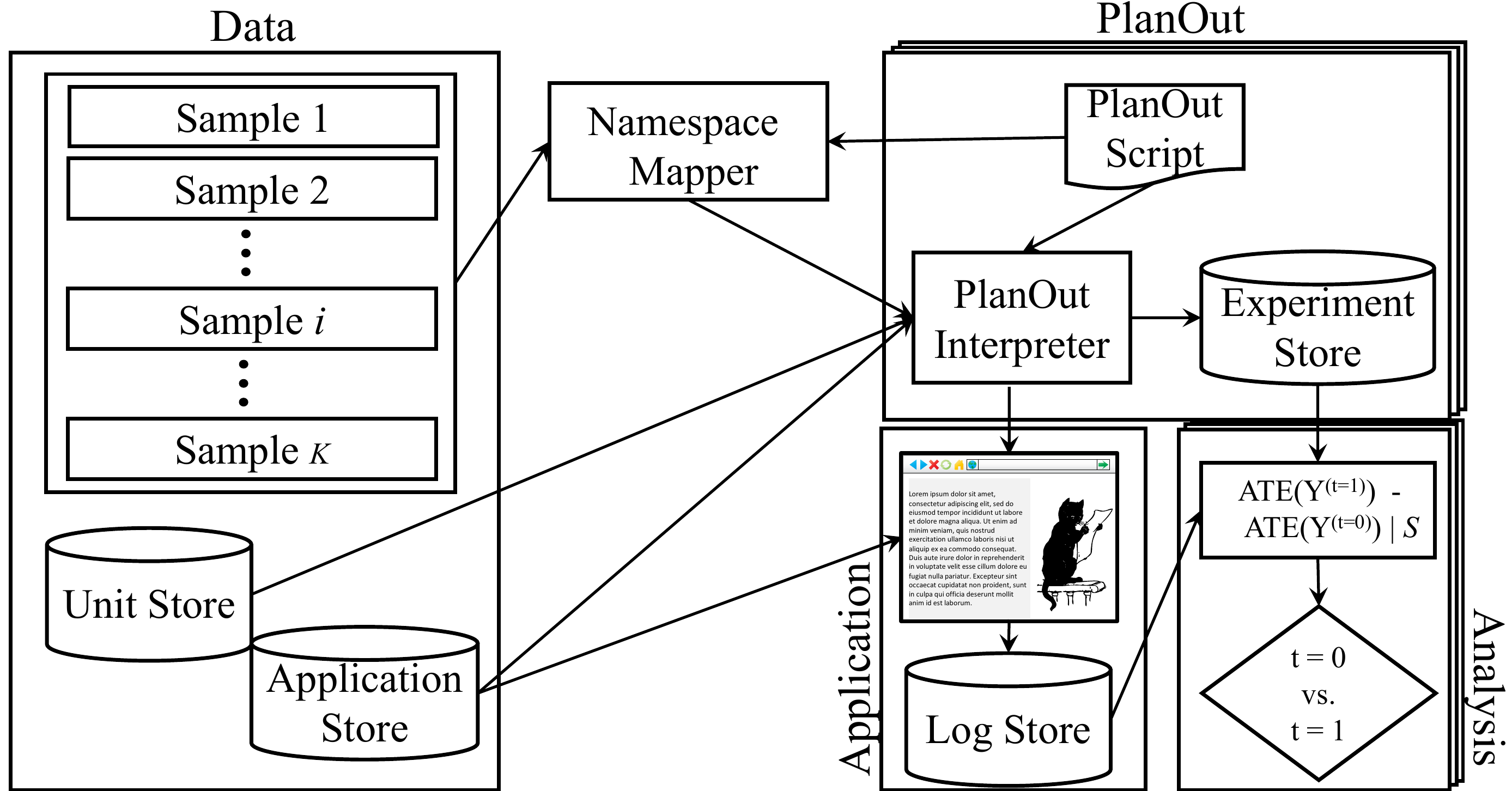}\\
    \caption{The structure of the \planout{} framework. A \emph{Namespace Mapper} randomly assigns a script to a randomly selected sample of the population of users. The \emph{\planout{} Interpreter} randomly assigns users to treatments, leveraging existing application infrastructure to record outcomes, as described in \S\ref{sec:online-experimentation}. The \emph{Analysis} box uses standard notation (e.g., \cite{gerber2012field}): $t$ represents a treatment instance from $T$, $Y$ represents the outcome variable, and $S$ represents the conditioning set. 
    \label{fig:experimentation-planout}}
\end{figure}
 Many firms have pre-existing methods for computing metrics of interest, and record these data regardless of the presence of an experiment. Figure~\ref{fig:experimentation-planout} depicts these existing systems in the ``Data'' and ``Application'' components of the diagram. The \planout{} framework can be added to record the assignment of units to treatments. Once the execution of a \planout{} script completes, there is a store that contains the mappings from variables defined in \planout{} to their values; these variables and their values are recorded in the ``Experiment Store.'' 

A \planout{} script specifies the assignment of treatments to units, although assignment does not happen until the \planout{} script is executed by the \planout{} interpreter. The interpreter takes as input: (1) the \planout{} program, (2) each unit $u$ of the sample of units specified in the selected sample(s), and (3) any unit-related data queried by external operators or application-specific data that might be required by the treatment procedure.  The interpreter deterministically hashes units to treatments in an effectively random fashion. 

Treatments and outcomes must be recorded in order to estimate causal effect. Outcomes typically correspond to variables or metrics already being recorded by the application's data-recording infrastructure. The \planout{} framework includes additional data-recording capabilities that record experiment metadata. 

\paragraph{Important Language Features} 
As a DSL built by domain experts, \planout{} implements functionality only relevant to experimentation. Consequently, \planout{} is not Turing complete: it lacks loops, recursion, and function definition. 
It has two control flow constructs (\lstinline{if/else} and \lstinline{return}) and a small core of built-in functions (e.g., \lstinline{weightedChoice}, \lstinline{bernoulliTrial}, and \lstinline{length}). 

On its surface, \planout{} may appear to share features with probabilistic programming languages (PPLs)~\cite{pfeffer2016practical, anglican, InferNET14, gordon2014tabular, gordon2014probabilistic}. PPLs completely describe the data generating process; in contrast, \planout{} programs specify only one part of the data generating process---how to randomly assign treatments---and this code is used to control aspects of a product or service that is the focus of experimentation. 

There are two critical features of \planout{} that differentiate it from related DSLs, such as PPLs: (1) the requirement that all random functions have an explicit unit of randomization, and (2) built-in control of data recording via the truth value of \planout{}'s \lstinline{return}.  Only named variables on paths that terminate in \lstinline{return true} are recorded. This is similar to the discarded executions in the implementation of conditional probabilities in PPLs. A major semantic difference between \planout{} and PPLs is that we expect \planout{} to have deterministic execution for an input. Variability in \planout{} arises from the population of inputs; variability in PPLs come from the execution of the program itself.

\paragraph{Framework System Assumptions} \planout{} abstracts over the sampling mechanism, providing an interface that randomly selects from pre-populated partitions of unit identifiers, corresponding to samples from the population of interest, as depicted on the far-left-hand side of Figure~\ref{fig:experimentation-planout}. The interface that selects samples is the \emph{Namespace Mapper}.  This component extracts the application parameters manipulated by a \planout{} script and hashes them, along with the current experiment name, to one or more samples.  We have spoken with data scientists and software engineers at several firms that use \planout{}, and they have stated that the mapping from experiments to samples was what drew them to the \planout{} framework.  The mapping avoids clashes between concurrently running experiments, which is one of the primary challenges of online experimentation~\cite{kohavi2009controlled, kohavi2013online}. Readers interested in the specifics of \planout{}'s hashing method for scaling concurrent experiments can refer to the paper~\cite{Bakshy:2014:DDO:2566486.2567967}; it is not relevant to \toolname{}'s analyses.

\section{\toolname{} Static Analysis Tool}
\label{sec:tool}
\toolname{} is a command-line tool written in OCaml that performs two main tasks. It: (1) checks whether the input script represents a randomized experiment by validating the presence of random assignment and the absence of any failures related to selection bias, unrecorded confounders, or positivity; and (2) generates all valid contrasts and their associated conditioning sets for the ATE estimator. 
\toolname{} translates \planout{} programs to an intermediate representation (\S\ref{sec:translate}) and assigns special labels to variables in the program (\S\ref{sec:qualifiers}). It then builds a data dependence graph (DDG), which it uses to generate contrasts for the ATE estimator (\S\ref{sec:ddg}).

\subsection{\planout{} Intermediate Representation}
\label{sec:translate}
Upon parsing, \toolname{} performs several routine program transformations. It: (1) converts variables to an identification scheme similar to SSA, (2) performs constant propagation, and (3) rewrites functions and relations (such as equality) in A-normal form~\cite{muchnick1997advanced, aho1986compilers, cytron1991efficiently, sabry1993reasoning}. Expressions may contain external function calls as subexpressions. Since it may not be possible to reason about the final values of a variable defined in a \planout{} program, \toolname{} reasons about intermediate values instead and reports results over a partially evaluated program~\cite{futamura1999partial}. 

After these routine transformations, \toolname{} splits the program into straight line code via tail duplication, such that every path through the program may be evaluated in isolation of the others. Although this transformation is exponential in the number of conditional branches, in practice the branching factor of \planout{} programs is quite small (discussed in \S\ref{sec:evaluation} and Figure~\ref{fig:efficiency}).

\begin{wrapfigure}{r}{0.25\textwidth}
\begin{lstlisting}[basicstyle=\incolcode]
if (max_br > 550) {
  // do something 
} else {
  // do something else
}
\end{lstlisting}
\caption{\label{fig:smt}\toolname{} leaves \texttt{max\_br} abstract.}
\end{wrapfigure}

\toolname{} then converts guards into assertions and uses the Z3 SMT solver to ensure that variables assigned along paths are consistent with these assertions~\cite{de2008z3}. For each assertion, \toolname{} queries Z3 twice---first to obtain a satisfying solution, and then to test whether this solution is unique. Evaluation of the intermediate representation may contain un-evaluated code, so if there is more than one solution, \toolname{} keeps the code chunk abstract. For example, the intermediate representation of code that branches on a threshold such as that depicted in Figure~\ref{fig:smt} would contain the lines \lstinline{fv1 = max_br > 550; assert fv1;}. \toolname{} would instantiate \lstinline{fv1}, but would keep \lstinline{max_br} abstract. When pretty printing the output as in Figure~\ref{fig:example-bitrate-analysis}, \toolname{} does not show intermediate values; the csv output,  however,  contains all intermediate variables and unevaluated expressions.

\toolname{} uses SSA and A-normal form because they aid in contrast generation: a single execution of a \planout{} program corresponds to the assignment of a unit to a treatment. However, additional intermediate variables can have somewhat ambiguous semantics when attempting to model a programmatically defined experiment causally; although they aid in e.g., the detection of causal sufficiency errors (\S\ref{sec:bugs}), they make reasoning about causal inference using tools such as causal graphical models quite difficult (\S\ref{sec:ddg}).

\subsection{Variable Labels for Causal Inference}
\label{sec:qualifiers}
The \planout{} language contains only some of the  necessary features for reasoning about the validity of experiments. Given only programs written in \planout{}, \toolname{} may not be able to reason about some common threats to internal validity. The interaction between random operators and control flow can cause variables to lose either their randomness or their variation. Furthermore, we need some way of guaranteeing that external operators do not introduce confounding. 

To expresses this missing information, we introduce a 4-tuple of variable labels ($rand$, $card$, $tv$, $corry$) that  \toolname{} attempts to infer and propagate for each \planout{} program it encounters~\cite{sabelfeld2003information, denning1976lattice}. Unsurprisingly, inference may be overly conservative for programs with many external functions or variables. To increase the scope of experiments \toolname{} can analyze, users may supply \toolname{} with global and local configuration files that specify labels.

\paragraph{Randomness ($rand$)}
\planout{} may be used with existing experimentation systems; this
means that there may already be sources of randomness available and
familiar to users. Furthermore, since \planout{} was designed to be
extensible, users may freely add new random operators. 
 
\paragraph{Cardinality ($card$)} The size of variables' domains (cardinality)
impacts an experiment's validity. Simple pseudo-random assignment requires high
cardinality units of randomization to properly balance the assignment
of units into conditions.  In the example program of Figure~\ref{code:example-bitrate-planout}, all variables have low cardinality, except for \lstinline{context}.

\paragraph{Time Variance ($tv$)} For the duration of a particular experiment, a given variable may be constant or time-varying. Clearly, some variables are always constant or always time varying.  For example, \emph{date-of-birth} is constant, while \emph{days-since-last-login} is time varying. However, there are many variables that cannot be globally categorized as either constant or time-varying. The $tv$ label allows experimenters to specify whether they expect a variable to be constant or time-varying over the duration of a given experiment. 

Since ATE/CATE assumes subjects receive only one treatment value for the duration of the experiment, \toolname{} cannot use them to estimate the causal effect of treatments or conditioning set variables having a \emph{tv} label. A \planout{} program may contain other valid contrasts assigned randomly, and independently from the time-varying contrasts; \toolname{} will still identify these treatments and their conditioning sets as eligible for being analyzed via ATE/CATE.

\textit{Example.} The first branch through the program in Figure~\ref{code:example-bitrate-planout} leads to assignments of \lstinline{max_bitrate} that vary with time. This part of the script encodes a contextual bandits experiment, an approach to experimentation that shares many features with reinforcement learning. Because bandits experiments use information from the environment in a loop to determine treatment, multiple visits to a website containing this sort of experiment could result in different treatments. Aggregating across individuals for a particular treatment (as ATE would do) is not sound in this case.

Since assignments along this branch cannot be compared in a between-subjects analysis, \toolname{} excludes them from the contrasts returned in Figure~\ref{fig:example-bitrate-analysis}; note that none of the contrasts compare across \lstinline{max_bitrate} when \lstinline{dynamic_policy} is set to true.

\paragraph{Covariates and Confounders ($corry$)} Many experiments use features of the unit to assign treatment (a type of \emph{covariate}; see Appendix~\ref{app:glossary}), which may introduce confounding. \toolname{} automatically marks external variables and the direct results of non-random external calls as correlated with outcome (i.e., $Y$). This signals that, if the variable is used for treatment assignment, either their values must be recorded, or sufficient downstream data must be recorded to recover their values. 

\textit{Example.} \toolname{} marks the variable \lstinline{country} in Figure~\ref{code:example-bitrate-planout} as correlated with $Y$. Although data from the \lstinline{country} variable affects treatment assignment, this information is captured by variables \lstinline{emerging_market} and \lstinline{established_market}, so \toolname{} raises no error.

\subsection{Data Dependence Graph (DDG)}
\label{sec:ddg}
\toolname{} builds a DDG to propagate variable label information ~\cite{ferrante1987program}.
Since \planout{} only has a single, global scope, its data dependence analysis is straightforward:
\begin{enumerate}[leftmargin=*]
	\item Assignment induces a directed edge from the references on the right-hand side to the variable name.
	\item Sequential assignment of $var_i$ and $var_{i+1}$ induces no dependencies between $var_i$ and $var_{i+1}$, unless the r-value of $var_{i+1}$ includes a reference to $var_i$.
	\item For an if-statement, \toolname{} adds an edge from each of the references in the guard to all assignments in the branches.
	\item In the case of an early return, \toolname{} adds edges from the variables in dependent guards to all variables defined after the return. 
\end{enumerate}

\begin{wrapfigure}[15]{r}[0pt]{0.6\textwidth}
\centering
\includegraphics[width=0.6\textwidth, draft=false]{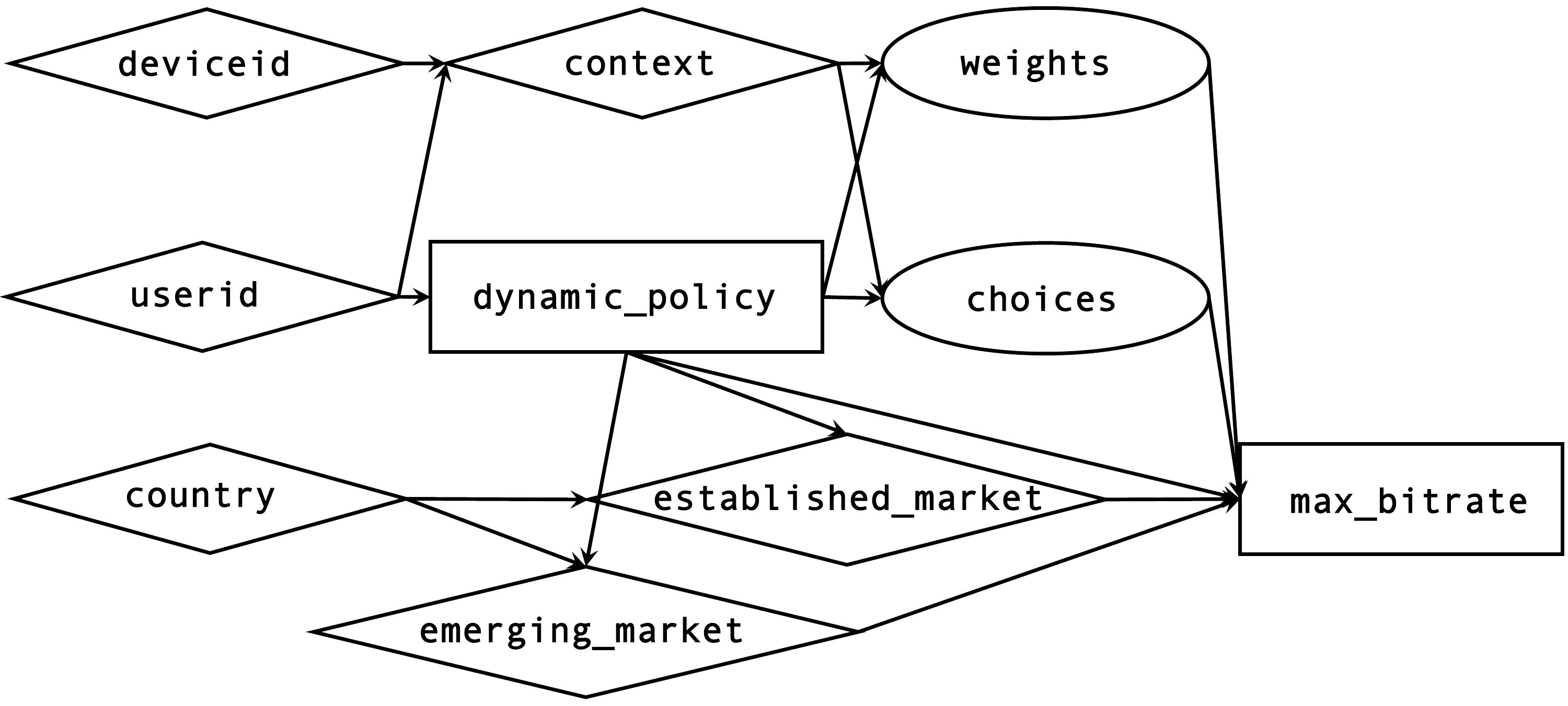}
\caption{\label{fig:example-bitrate-ddg} The DDG that \toolname{} produces for the \planout{} program in Figure~\ref{code:example-bitrate-planout}. Diamonds denote non-random external variables and their dependents. Rectangles denote random variables. \toolname{} does not consider variables whose \emph{existence} depends on a random variable, but have no variation in their values, to be random.}
\end{wrapfigure}

Random, independent assignment implies independence between potential causes, so long as the (possibly empty) conditioning set has been identified and recorded. \toolname{} computes the DDG for the full script and uses the full DDG to determine when it is possible to to marginalize over some variables. 
Figure~\ref{fig:example-bitrate-ddg} shows an example DDG for the example experiment of Figure~\ref{code:example-bitrate-planout}.

\paragraph{Propagating Variable Labels.} \toolname{} marks variables directly assigned by built-in random functions or external random functions as random. The randomness label takes a tuple of identifiers as its argument. This tuple denotes the unit(s) of randomization, used for reasoning about causal estimators. Any node with a random ancestor is marked as random (with the exception of variables that do not vary), with units of randomization corresponding to the union of the ancestors' units. \lstinline{max-bitrate} in Figure~\ref{fig:example-bitrate-ddg} is random, even though it is set to a constant in Figure~\ref{code:example-bitrate-planout}. This is because assignment is still randomly assigned on the basis of \lstinline{dynamic_policy}.

If a random operator uses a low-cardinality unit of randomization, it will be marked as non-random. Note, however, that if the unit of randomization for a random function is a tuple with at least one high cardinality variable, then the resulting variable will remain random.

\toolname{} propagates time-varying labels in the same manner as random labels. Unlike randomness, there is no interaction between the time-varying label and any other label.

\paragraph{Converting DDGs to Causal Graphical Models (CGMs)}
Readers familiar with graphical models may wonder whether the DDG can be transformed into a directed graphical model. Programmatically defined experiments have two features that, depending on context, make such a transformation either totally inappropriate or difficult to extract: (1) deterministic dependence; and (2) conditional branching. These two features can induce what's known as ``context-sensitive independence,'' which limits the effectiveness of existing algorithms that would otherwise make graphical models an appealing target semantics. Although some work has sought to remedy branching, treatment of context-sensitive independence in graphical models more broadly is an open research problem~\cite{minka2009gates}. Furthermore, from a practical perspective, it is unclear how the versioned variables in the DDG ought to be unified, and some variables simply don't belong in a CGM (e.g., \lstinline{userid}).

\section{Detecting Threats to Internal Validity}
\label{sec:bugs}
We characterize some static threats to internal validity based on the forms of bias in experimental design, described in \S\ref{sec:background}.
Note that because there is currently no underlying formalism for the correctness of online field experiments that maps cleanly to a programming language context, we cannot define a soundness theorem for programmatically defined experiments. Some of the threats described below would be more properly considered code smells, rather than outright errors~\cite{fowler2018refactoring}.

\paragraph{Randomization Failures}

There are three ways a \planout{} program may contain a failure of
randomization: (1) when it records data along a path that is not randomized, (2) when the units of randomization have low cardinality, and (3) when it
encounters path-induced determinism. 

Recording data along non-randomized paths occurs when
there exists at least one recorded path through the program that \emph{is}
randomized and at least one recorded path through the program that is \emph{not} 
randomized. Imagine the engineering team from Section~\ref{sec:introduction} ran only the country-level design  (i.e., Lines 9--17 of Figure~\ref{code:example-bitrate-planout}). Then, data for users from outside the four countries of interest would be recorded in the experiment store in Figure~\ref{fig:experimentation-planout}, alongside data for users from within those countries. \toolname{} raises an error for this path. The fix is simple: add \lstinline{return false} after Line 15. 

Units of randomization must have significantly higher cardinality than experimental
treatments to ensure that each treatment is assigned sufficient
experimental units to make valid statistical inferences about the population. Users can correct this by either annotating the unit of randomization as having high cardinality, or re-assessing their choice of unit. 

Data-flow failures of randomization occur when inappropriate computations flow into units. \planout{} allows units to be the result of arbitrary computations: e.g., one example \planout{} script in the corpus described in Section~\ref{sec:corpus} sets the unit of randomization to be \lstinline{userid * 2}. A \planout{} user might want to do this when re-running an experiment, to ensure that at least some users are assigned to a new treatment. However, this feature can lead to deterministic assignment when used improperly. The following is a syntactically valid \planout{} program; \toolname{} detects an error in it when it converts the program to its intermediate representation:

\begin{center}
\begin{minipage}{0.6\linewidth}
\begin{lstlisting}[basicstyle=\incolcode]
max_br = uniformChoice(choices=[400, 900], unit=userid);
dynamic_policy = bernoulliTrial(p=0.3, unit=max_br);
\end{lstlisting}  
\end{minipage}
\end{center}
\noindent When writing this code, the researcher may believe that there are four possible assignments for the pair of variables. However, because the assignment of input units to a particular value is the result of a deterministic hashing function, every user who is assigned \lstinline{max_br=400}, is assigned the same value of \lstinline{dynamic_policy} because the input to the hash function for \lstinline{bernoulliTrial} is always 400. Therefore, they will never record both (400, true) and (400, false) in the data, which likely contradicts the programmer's intent. 

\paragraph{Treatment Assignment Failures} 
\toolname{} requires that all assigned treatments along a path have the possibility of being assigned to at least one unit, and that at least some treatments may be compared. There are three ways a \planout{} program may contain a failure of treatment assignment, when: (1) some treatment has a zero probability of being assigned (i.e., a positivity error); (2) there are fewer than two treatments that may be compared along a path; and (3) dead code blocks containing treatment assignment.

Syntactically correct \planout{} code permits
users to set probabilities or weights to zero, either directly or as the result of evaluation. A zero-valued weight may flow in
from earlier computation or be due to type puns or conversions. Furthermore, to establish a causal relationship between variables, there must be at least two
alternative treatments under comparison. 

When \toolname{} expands
all possible worlds to generate all possible treatments, it checks that there is at least one assignment for
the free variables in the guard that causes the guard to evaluate to true and
that there is at least one assignment that causes it to evaluated to false. If
only one final value (i.e., true or false) is possible, then \toolname{} raises
an error. 

\paragraph{Causal Sufficiency Errors} One of the main assumptions underlying causal reasoning is \emph{causal sufficiency}, or the assumption that there are no unmeasured confounders in the estimate of treatment effect. 
Barring run-time failures, we have a complete picture of the assignment mechanism in \planout{} programs. Unfortunately, a \planout{} program may allow an unrecorded variable to bias treatment assignment. 

Consider a program that assigns treatment on the basis of user country, accessed via a \texttt{get\-User\-Country} function:
\begin{center}
\begin{minipage}{0.65\linewidth}
\begin{lstlisting}[basicstyle=\incolcode]
if (getUserCountry(userid=userid) == 'CA') {
  mxbr = uniformChoice(choices=[75, 90], unit=userid);
} else {
  mxbr = uniformChoice(choices=[40, 75, 90], unit=userid);
}
\end{lstlisting}
\end{minipage}
\end{center}
Treatment assignment of \lstinline{mxbr} depends on user country, so user country is a potential confounder. \toolname{} will convert the guard to A-normal form and assign the \lstinline{getUserCountry} function call to a fresh variable. Because this variable does not appear in the input program text, it cannot be recorded by the \planout{} framework's data recording system. Therefore, the program and resulting analyses will violate the causal sufficiency assumption.

If \toolname{}
encounters a static error or threat, it reports that the script failed to pass
validation and gives a reason to the user.  Some of the fixes are easy to
determine from the error and could be interpolated automatically. We leave this to future work.  Other errors
require a more sophisticated understanding of the experiment the script represents and can only be determined by the script's author. 

\section{\planout{} Corpora}
\label{sec:corpus}
\definecolor{mygray}{gray}{0.85}
\definecolor{lightgray}{gray}{0.95}
\begin{table}
\small\caption{\label{tab:corpus}Descriptive statistics for the corpus. \emph{Input variables} correspond to the number of unique variables in the input program. \emph{IR paths} correspond to the number of paths generated for the intermediate representation of the program. The unique authors listed refer to authors that only appear in that corpus. There was overlap in authorship between each of the corpora, with 15 authors appearing in all three. The data was collected over a two year period.}

\begin{tabular}{| *{4}{p{0.15\columnwidth} r|}}
\hline
\rowcolor{mygray}
\multicolumn{8}{|c|}{Corpus \planout{}-A: Confirmed Deployed and Contrasts Analyzed}\\
\hline
\multicolumn{4}{|l}{Unique \planout{} Scripts} & \multicolumn{4}{r|}{566} \\
\multicolumn{4}{|l}{Unique Experiments} & \multicolumn{4}{r|}{240} \\
\multicolumn{4}{|l}{Unique Authors (Total)} & \multicolumn{4}{r|}{30 (70)}\\
\hline
Min. Versions & 1&   Min. LOC & 1 & Min. Input Vars. & 1 & Min. IR Paths & 1 \\
Med. Versions & 3&   Med. LOC & 45& Med. Input Vars. & 9 & Med. IR Paths & 4\\
Avg. Versions & 4&  Avg. LOC & 78& Avg. Input Vars. & 10 & Avg. IR Paths & 29\\
Max. Versions & 28& Max. LOC & 691& Max. Input Vars. & 60 & Max. IR Paths &  6561\\
\hline
\rowcolor{mygray}
\multicolumn{8}{|c|}{Corpus \planout{}-B: Confirmed Deployed and Contrasts Recorded}\\
\hline 
\multicolumn{4}{|l}{Unique \planout{} Scripts} & \multicolumn{4}{r|}{381} \\
\multicolumn{4}{|l}{Unique Experiments}        & \multicolumn{4}{r|}{130} \\
\multicolumn{4}{|l}{Unique Authors (Total)}            & \multicolumn{4}{r|}{25 (72)}\\
\hline
Min. Versions & 1 &  Min. LOC & 1 & Min. Input Vars. & 1 & Min. IR Paths & 1 \\
Med. Versions & 3 & Med. LOC & 26& Med. Input Vars. & 7 & Med. IR Paths &  3\\
Avg. Versions & 4 & Avg. LOC & 41 & Avg. Input Vars. & 7 & Avg. IR Paths & 7 \\
Max. Versions & 32 & Max. LOC & 495 & Max. Input Vars. & 26 & Max. IR Paths & 253 \\
\hline
\rowcolor{mygray}
\multicolumn{8}{|c|}{Corpus \planout{}-C: Not Deployed}\\
\hline 
\multicolumn{4}{|l}{Unique \planout{} Scripts} & \multicolumn{4}{r|}{493} \\
\multicolumn{4}{|l}{Unique Experiments}        & \multicolumn{4}{r|}{74} \\
\multicolumn{4}{|l}{Unique Authors (Total)}            & \multicolumn{4}{r|}{23 (47)}\\
\hline
Min. Versions & 1&  Min. LOC & 1 & Min. Input Vars. & 1 & Min. IR Paths & 1o\\
Med. Versions & 3 & Med. LOC & 56& Med. Input Vars. & 13 & Med. IR Paths &  4\\
Avg. Versions & 8 & Avg. LOC & 137 & Avg. Input Vars. & 16 & Avg. IR Paths & 368\\
Max. Versions & 124 & Max. LOC & 883 & Max. Input Vars. & 48 & Max. IR Paths & 27675\\
\hline
\end{tabular}
\end{table}

\companyname{} provided a corpus of \planout{} scripts that we used to evaluate \toolname{} via a single point of contact. This corpus contains every \planout{} script written between 3 August 2015 and 3 August 2017. The actual dates (i.e., the choice of 3 August) was arbitrary.  The start year was at a late enough point after \planout{}'s introduction at Facebook that the language implementation was stable. The timeframe allowed us to be sure that the experiments in the corpus were completed at the time of analysis. Facebook also provided us with a corpus of manually specified contrasts that were used in the analysis of the experimentation scripts that were actually deployed. 

Each experiment may have been updated in vivo or may have a temporary (but syntactically valid) representation captured by a snapshotting system, leading to multiple versions of a single experiment. Some experiments are programmatically generated, leading to verbose experiments that are much longer than what a human might write.

The tool used for analyzing scripts can only be used for ATE analysis (not CATE), and so it provides a meaningful point of comparison for \toolname{}. While we do not have access to the custom analyses of more complex experiments (e.g., database queries, \textit{R} code, etc.) we can infer some characteristics of the intended analysis by partitioning the corpus into three sub-corpora:
\begin{description}[leftmargin=1em]
    \item[\textbf{\planout{}-A}] This corpus contains scripts that were analyzed using some form of ATE (i.e., $Avg(Y|T_1=t^0_i, \ldots T_n=t^0_n)- Avg(Y|T_1=t^1_i, \ldots T_n=t^1_n)$), where the variables $T_1,\ldots T_n$ were manually specified and automatically recorded during the duration of the experiment. Users may manually specify that a subset of the recorded variables be continuously monitored for pairwise ATE. 
    Neither the recording, nor the data analysis tools have any knowledge of \planout{}. This is the main corpus we will use for evaluating \toolname{}, since the goal of \toolname{} is to automate analyses that firms such as \companyname{} must now do manually.
    \item[\textbf{\planout{}-B}] Some scripts have data recorded, but no automated analyses. This may be because the scripts are not suited to ATE. We analyze the scripts in this corpus to see whether there are any CATE analyses that \toolname{} can identify.  This corpus may also contain custom analyses for within-subjects experiments, contextual bandits experiments, experiments that must account for peer effects, etc. 
    \item[\textbf{\planout{}-C}] These are scripts that have never been deployed and therefore may not have had the oversight of domain experts. This corpus provides the best approximation of the kinds of mistakes that \planout{} users actually make. 
\end{description}

Note that users at \companyname{} are typically either experts in the domain of the hypotheses being tested or they are analysts working directly with domain experts. Therefore, \planout{}-C was our best chance of finding scripts by non-experts in experimental design (although they were likely still domain experts). In all cases, experiments undergo review before being deployed. Table~\ref{tab:corpus} gives a more detailed description of the corpora.

\subsection{Characterizing Representative \planout{} Programs} 
\label{sec:representative-programs}
\begin{figure}
    \centering
    \begin{minipage}{0.45\textwidth}
    \begin{lstlisting}[numbers=left, basicstyle=\incolcode, xleftmargin=2.5em]
show_feature = true;
in_exp = false;
if (userid == 123456789) {
  in_exp = true;
  if (post_has_photo == true) {
    show_feature = false;
  } 
  if (post_has_video == true) {
    show_feature = false;
  } 
  return true;
} 
in_pop = extPred(ep="in_pop", userid=userid);
if (!in_pop || userid == 0) {
  return false;
} 
in_exp = bernoulliTrial(p=1/100, unit=post_id);
if (!in_exp) {
  return false;
} 
if (post_has_photo == true) {
  show_feature = false;
} 
\end{lstlisting}
    \end{minipage}\quad
\begin{minipage}{0.45\textwidth}
    \begin{lstlisting}[numbers=left, basicstyle=\incolcode, xleftmargin=2.5em]
skip_logging = true;
enabled = false;
if (extPred(userid=userid, ep="feat_exp")) {
  test = bernoulliTrial(p=1/25, unit=userid);
  if (test) {
    enabled = bernoulliTrial(p=1/2, unit=userid);
    skip_logging = false;
  } else {
    enabled = true;
  }
} else if(extPred(userid=userid, ep="feat_rollout")) {
  enabled = true;
}
if (enabled) {
  animation_type = "WINDOW_ANIMATION";
} else  {
  animation_type = "CONTROL";
}
if (skip_logging) {
  return false;
}
    \end{lstlisting}
\end{minipage}
\caption{\label{fig:example-programs}Some representative, lightly edited and anonymized experiments written in \planout{}. \textit{Left:} This script mixes testing code with experimentation code. Lines 5-12 set values for user 123456789 and record those values. The actual experiment is in lines 14-26. It is only conducted on the population defined by the external predicate and the user being recorded in (represented here when the userid is 0). \toolname{} raises an error for this script. \textit{Right}:   \toolname{} raises a Causal Sufficiency error for this lightly edited and anonymized experiment, assuming that the call to \lstinline{extPred} in line 3 is non-random.}
\end{figure}

We designed \toolname{}'s analyses on the basis of the universe of syntactically valid \planout{} programs and our domain knowledge of experimentation. We built \toolname{} from the perspective that (1) \planout{} is the primary means by which experimenters design and deploy experiments, but (2) they can use other systems, if they exist. \companyname{} uses many experimentation systems and has a variety of human and code-review methods for the functionality that \toolname{} provides. Therefore, we wanted to know: what are some characteristics of \planout{} programs that people actually write and deploy?

We found that analysts at \companyname{} used \planout{} in a variety of surprising ways and had coding habits that were perhaps indicative of heterogeneity in the programming experience of authors. 
Through conversations with \companyname{}, we have come to understand that most \planout{} users can be described along the two axes depicted in Figure~\ref{fig:experience}.

Table~\ref{tab:error_counts} enumerates the errors raised by \toolname{} over the three corpora. 
Each warning does not necessarily indicate an error during deployment or analysis, due to the fact that there are pre-existing mechanisms and idiosyncratic usages of \planout{}.

Sections~\ref{sec:characteristics-planout-a}--\ref{sec:characteristics-planout-c} give detailed descriptions of the characteristics we observed in the three \planout{} corpora, while Section~\ref{sec:rq1-findings} summarizes our findings.

\begin{wrapfigure}{R}{0.5\columnwidth}
    \centering
    \begin{tabular}{*{2}{l}|*{2}{>{\centering\arraybackslash}m{0.1\columnwidth}|}}
        \multicolumn{2}{c}{} & \multicolumn{2}{c}{Programming Experience}\\
        \multicolumn{2}{c}{} & \multicolumn{1}{c}{\small{High}} & \multicolumn{1}{c}{\small{Low}}\\
        \cline{3-4} 
        \multirow{3}{*}{\rotatebox{90}{\parbox{1cm}{Experimental\\Design\\Experience}}} & \rotatebox{90}{\small{High}} & I & II \\
        \cline{3-4}
        & \rotatebox{90}{\small{Low}} & III & IV\\
        \cline{3-4}
    \end{tabular}
    \caption{\label{fig:experience}Experience matrix for \planout{} authors. The horizontal axis represents programming experience or ability; the vertical axis represents experience in experimental design. We believe most authors represented in the \planout{} corpora are in quadrants I and II.  \toolname{}'s novel analyses target experiment authors in quadrants I-III and may be especially useful for authors in III, whom we believe are under-represented in the corpora. We conjecture, but cannot verify, that most of the errors \toolname{} flags in the corpora belong to authors in II.}
\end{wrapfigure}

\newcounter{rowcntr}[table]
\renewcommand{\therowcntr}{\arabic{rowcntr}}

\newcolumntype{O}{>{\textbf{OC\refstepcounter{rowcntr}\therowcntr}:~}l}
\AtBeginEnvironment{tabular}{\setcounter{rowcntr}{0}}
\begin{table}
\small\caption{\label{tab:error_counts}The counts of code smells, static script errors, and tool failures found when running \toolname{} on the corpora. A \toolname{} error does not necessarily indicate that the experiment was run in error. A single experiment may have many script versions, not all of which were deployed. The numbers for \planout{}-A reflect the state of the corpus after adjustments for easily fixed type inconsistencies (initially 87), since we know those scripts ran in production, and wanted to see if \toolname{} could find more interesting errors or smells. There were no adjustments to the other two corpora. The grey band represents manual analysis, only performed on \planout{}-A. \textbf{*}We analyzed \planout{}-A with a max number of random variable choices of 100 and the other two corpora with the default setting of 20.}
\begin{tabular}{|l|*{3}{rr|}}
    \hline
  \rowcolor{mygray}
  \multicolumn{1}{|c}{}& \multicolumn{2}{c}{\planout{}-A} & \multicolumn{2}{c}{\planout{}-B} & \multicolumn{2}{c|}{\planout{}-C}\\
  \rowcolor{mygray}
  \multicolumn{1}{|c}{} & Scripts & \multicolumn{1}{c}{Exps.} & Scripts & \multicolumn{1}{c}{Exps. } & Scripts & \multicolumn{1}{c|}{Exps. } \\
    \rowcolor{mygray}
  \multicolumn{1}{|c}{Output Category} &(566) & \multicolumn{1}{c}{(240)} &  (381) & \multicolumn{1}{c}{(130)} &  (493) & \multicolumn{1}{c|}{ (74)} \\
  \hline 
  \label{oc:not-experiment}Not an experiment & 10 & 10 & 8 & 5 & 22 & 8\\
  \label{oc:unit-low-cardinality}Low cardinality unit & 7 & 1 & 6 & 2 & 1 & 1\\
  \label{oc:ambiguous-semantics}Ambiguous semantics & 5 & 2 & 0& 0& 0&0\\
  \label{oc:type-inconsistencies}Type inconsistencies & 10 & 4 & 36 & 12 & 4 & 2\\
  \label{oc:causal-sufficiency}Causal sufficiency errors & 111 & 54 & 75 & 22 & 77 & 17 \\
  \rowcolor{lightgray}
  \multicolumn{1}{|l|}{\quad False positive} & 47 & 23& & & &\\
  \rowcolor{lightgray}
  \multicolumn{1}{|l|}{\quad Testing code} & 23 & 8& & & &\\
  \rowcolor{lightgray}
  \multicolumn{1}{|l|}{\quad Possible random assignment}& 41& 23& & &&\\
  \label{oc:no-randomization}Recorded no randomization & 25 & 11 & 83 & 23 & 214 & 35\\
  \rowcolor{lightgray}
  \multicolumn{1}{|l|}{\quad Missed paths (tests)} &4 & 1 & & & &\\
  \rowcolor{lightgray}
  \multicolumn{1}{|l|}{\quad No randomization (config)} & 12 & 7 & &&& \\  
  \rowcolor{lightgray}
  \multicolumn{1}{|l|}{\quad Possible random assignment} & 9 & 3& &&&\\
  \label{oc:rv-no-variation}Random variable no variation & 2 &2& 22 & 15 &0&0\\
  \label{oc:too-many-choices}Exceeds max choices\textbf{*} & 0 & 0 & 21 & 14 & 0 & 0\\
  \label{oc:no-positivity}No positivity & 7 & 3 & 0 & 0 & 9 & 4\\
  \label{oc:dead-code}Dead code &  5 & 4 & 1 & 1 & 4 & 2\\
  \label{oc:not-implemented}Feature not implemented in tool & 29 & 8 & 29  & 10 &37  &7\\
  \label{oc:stack-overflow}Overflow error & 0 & 0 & 0 & 0 & 80 & 4 \\
\hline 
\end{tabular}
\end{table}

\subsubsection{Characteristics of \planout{}-A} 
\label{sec:characteristics-planout-a}
\planout{}-A contains our gold-standard data: all scripts were vetted by experts before deployment, with some component analyzed using ATE. Figure~\ref{fig:example-programs} provides some lightly anonymized example programs that \toolname{} identified as having potential errors. Their style and structure are good representations of real-world \planout{} programs.

\paragraph{Ambiguous Semantics and Type Errors} Since \toolname{} must initially perform type inference, it found 87 scripts in \planout{}-A that had typing errors. By far the most common issue flagged was the treatment of 0 as falsey. Upon manually inspecting the scripts, we found that most scripts could be modified so that these variables were consistently Boolean or numeric, depending on usage. Other typing issues included string values such as \lstinline{"default"} and \lstinline{"status_quo"} for numeric variables and guards such as \lstinline{userid == 0 || userid == "0"}, which suggest there might be some utility in providing our type checking facility to users of \planout{}. 

We also found three scripts from one experiment that applied the modulus operator to a fraction; since \planout{} uses the semantics of its enclosing environment for numeric computation, this script will return different values if it is run using languages with different semantics for modulus, such as PHP versus JavaScript. 

\paragraph{Modifying Deployment Settings within Experimentation Logic.} Some of the scripts marked as not experiments begin with \lstinline{return false} and had an unreachable and fully specified experiment below the return statement. \toolname{} flags dead code in \planout{} programs, since it can be the result of a randomly assigned variable causing unintended downstream control flow behavior. However, every dead code example we found had the form 
\lstinline{condition = false; if (condition) ...}
These features occurred exclusively in experiments that had multiple scripts associated with them that did not raise these errors. After discussing our findings with \companyname{}, we believe that this might be a case of \planout{} authors modifying the experiment while it is running to control deployment, rather than leaving dead-code in by accident, as it appears from \toolname{}'s perspective.

\paragraph{Using \planout{} for Application Configuration.}
One of the most surprising characteristics we found in \planout{}-A was the prevalence of using \planout{} for application configuration, \`a la Akamai's ACMS system or Facebook's Gatekeeper~\cite{sherman2005acms, tang2015holistic}. When these scripts set variables, but properly turned off data recording (i.e., returned false), \toolname{} marked them as not being experiments. When they did not turn off logging, they were marked as recording paths without randomization. Some instances of application configuration involved setting the support of a randomly assigned variable to a constant or setting a weight to zero. Since experiments require variation for comparison, \toolname{} raises an error if the user attempts to randomly select from a set of fewer than two choices. Three scripts contained expressions of the form \lstinline{uniformChoice (choices=[v], unit=userid)} for some constant value $v$.

As a result, users who aim to use \planout{} as a configuration system have no need for \toolname{}, but anyone writing experiments would consider these scripts buggy.

\paragraph{Mixing External Calls to Other Experimentation Systems} Almost 20\% of the scripts (106) include calls to external experimentation systems. In a small number of cases, \planout{} is used exclusively for managing these other systems, with no calls to its built-in random operators.

\paragraph{Non-read-only Units.} One of the other firms we spoke to that uses \planout{} treats units of randomization as read-only, unlike other variables in \planout{} programs. \companyname{} does not do this. Therefore, programs that manipulate the unit of randomization may be legal: for instance, the aforementioned instance where the unit was set to \lstinline{userid * 2}. We also observed a case where the unit was set to be the result of an external call---without knowing the behavior of this external call it is assumed to be low cardinality. In this case, the experiment was performing cluster random assignment, which is not covered by ATE and out of scope for \toolname{}.

\paragraph{Inspection for CATE} We investigated the conditioning sets \toolname{} produces to see whether there were any contrasts that ought to be computing CATE, rather than ATE. Unfortunately, our gold standard annotation set \planout{}-A does not include any experiments that were analyzed with CATE, as the labels were collected with a system that only supports ATE. \toolname{} produced conditioning sets for seven experiments; warning that ATE would not be valid for certain subsets of variables, but the gold truth data told us these experiments were aimed at learning variables for which ATE was valid (i.e., they were similar to the comparison between policies listed first in Fig~\ref{fig:example-bitrate-analysis}). Thus, these seven experiments produced false alarms. However, it is reasonable for \toolname{} to produce these conditioning sets because it does not have access to the variables of interest through the \planout{} language.

\paragraph{Tool Limitations} Twenty-two scripts required some more complex transformations to SMT logic that we have not yet implemented: all cases involved reasoning about map lookups or null values. The remaining seven scripts were all versions of a single experiment that used the \lstinline{sample} function, which \toolname{} does not currently support; this function generates $\binom{n}{k}$ subsets of size $k$ from a list of size $n$, but was left for future work since it is so rarely used. 

We did not expect to see any real causal sufficiency errors, due to the expert nature of the authors of \planout{}-A. Rather, we expect to see some false positives, due to the fact that \toolname{} is aggressive about flagging potential causal sufficiency errors. We made this design choice because the cost of unrecorded confounders can be very high. Furthermore, the fix is quite easy and can be automated, were \toolname{} to be integrated in a \planout{} editor. The errors that were not false positives were either cases in which the author intermingled testing code with experiment code, or where branching depending on an external function call that is sometimes random.

As an example, the code example on the left side of Figure~\ref{fig:example-programs} contains testing code mixed with experiment code. This scripts raises an error due to the fact that there is a recorded path with no randomization on it. The right side of Figure~\ref{fig:example-programs} is an example of a causal sufficiency error; the result of the call to \lstinline{extPred} is not recorded, but could be correlated with both treatment and outcome.

\subsubsection{Characteristics of \planout{}-B} 
\label{sec:characteristics-planout-b}
Our main interest in \planout{}-B is to identify whether there are experiments that could benefit from CATE analysis. Note that in \S\ref{sec:characteristics-planout-a}, we found several experiments that were eligible to be analyzed with both ATE and CATE. In \planout{}-B we found 14 scripts spanning nine experiments that contain analyses eligible for CATE.

\planout{}-B, as a corpus, has very similar characteristics to \planout{}-A: authors still mix deployment logic with experimentation logic, use \planout{} for what appears to application configuration, and use idiomatic expressions that may not typecheck. 

\subsubsection{Characteristics of \planout{}-C}
\label{sec:characteristics-planout-c}
Recall that scripts in \planout{}-C were never deployed. Investigating a subset of these scripts, we believe that this corpus is largely filled with scripts trying out certain features. For example, one extremely large script appears to be automatically generated. \planout{}-C was the only corpus that caused the tool to crash. As depicted in Table~\ref{tab:corpus}, the maximum number of paths in this corpus is an order of magnitude more than \planout{}-A. Figure~\ref{fig:efficiency}, which we discuss in depth in~\ref{sec:performance}, depicts how this very large number of paths contributes to the runtime. 

\subsubsection{Findings}
\label{sec:rq1-findings}
\planout{} scripts in deployment at \companyname{} represent a range of experimental designs. We observed factorial designs, conditional assignment, within-subjects experiments, cluster random assignment, and bandits experiments in the scripts we examined. 

\planout{} has the look and feel of writing Python, R, or other scripting languages popular among data scientists. However, without a unified coding style, and no restrictions on program correctness other than parsing, there is considerable variability in the ways experiment authors use \planout{}. This variability includes implementing behavior that \planout{} is not suited to solve. 

\section{Evaluation}
\label{sec:evaluation}
Real-world \planout{} scripts unsurprisingly contained few errors, since they were primarily written and overseen by experts in experimental design. For example, of the 25 recorded paths with no randomization, nine contained a special gating function that may sometimes be random, depending on its arguments. Four of the scripts appeared to be using \planout{} for configuration, leaving twelve scripts that essentially implemented application configuration logic. 
\renewcommand{\arraystretch}{1.5}
\begin{table}[t]
    \centering
    \small\caption{Mutations injected into sample \planout{} programs.}
    \begin{tabular}{lp{0.45\columnwidth}p{0.35\columnwidth}}
    \rowcolor{mygray}
         \multicolumn{1}{c}{Mutation} &  \multicolumn{1}{c}{Description} & \multicolumn{1}{c}{Fault?} \\
         \hline
         SAI & Subpopulation Analysis Insert; wraps a node of the AST in an if-then-else state where the guard is a feature of population, such that if the guard is true, the wrapped node is executed and recorded, and if the guard is false, the program returns false.  & \textit{Never}: creates a subpopulation frame that should \emph{not} be in the conditioning set. \\
         \cline{2-3}
         CI & Constant Insert; inserts a variable assignment from a constant. & \textit{Never}: constant assignment should have no bearing on the presence of errors. They may sometimes affect the treatments or conditioning sets.\\
         \cline{2-3}
         EFCI &  External Function Call Insert; inserts a variable assign\-ment from an external function call that could be correlated with outcome. & \textit{Never}: the variable defined cannot be correlated with treatment assignment.\\
         \cline{2-3}
        RSI & Return Statement Insert; inserts a return statement at an arbitrary point in a statement tree. The only restriction is that a new return statement cannot be added after another return statement. 
           & \textit{Sometimes}; a return true statement may be inserted before a return false, causing an error.\\
           \cline{2-3}
         CSE & Causal Sufficiency Error; wraps a node of the AST in an if-then-else statement where the guard is a feature of the population. & \textit{Sometimes}; if the mutation induces a dependency between the guard and treatment assignment, then the script will contain an error.  \\
         \cline{2-3}
         URE-1 & Unit of Randomization Error; replaces a unit of randomization with an expression containing the former unit and constants. & \textit{Sometimes}: some operations can reduce the cardinality of the unit, for example, modulus.\\
         \cline{2-3}
         URE-2 & Unit of Randomization Error; replaces the unit with another high cardinality unit. & \textit{Sometimes}: replacing the unit may make some treatments within-subjects.\\
         \cline{2-3}
         URE-3 & Unit of Randomization Error; replaces the unit with another variable defined previously in the program & \textit{Sometimes}; variables defined in the program should almost always have low-cardinality, however sometimes authors include functions of e.g. userid.
    \end{tabular}
    \label{tab:mutations}
\end{table}

Therefore, to test how well \toolname{} finds errors, we selected a subset of fifty scripts from \planout{}-A and mutated them. Table~\ref{tab:mutations} describes the mutations we performed and the type of effect on output we expected. 
We then validated a subset of the contrasts \toolname{} produced against a corpus of hand-selected contrasts monitored and compared by an automated tool used at \companyname{}. Finally, we report on \toolname{}'s performance,  since its effectiveness requires accurately identifying meaningful contrasts within a reasonable amount of time.

\renewcommand{\arraystretch}{1}
\begin{table}
\caption{We apply each type of mutation at a rate proportional to the eligible nodes in the input program's AST. We found the overall rates by applying our mutations over the \planout{}-A corpus. We assessed \toolname{} on over fifty randomly selected scripts from \planout{}-A. \toolname{} had a precision and recall of both 92\%. We these outcomes in \S\ref{sec:rq2-findings}.}
    \begin{subtable}{0.3\columnwidth}
    \centering
        \caption{\label{tab:mut-freq} Mutation proportions.}
    \begin{tabular}{|lc|}
        \hline
        CI      &0.22\\
        CSE     &0.20\\
        EFCI    &0.18\\
        RTI     &0.15\\
        SAI     &0.22\\
        URE     &0.02\\
        \hline
    \end{tabular}
    \end{subtable}
    \begin{subtable}{0.6\columnwidth}
        \caption{\label{tab:mutation-results}Mutation results.}
    \centering
    \begin{tabular}{l|*{4}{c}|}
    \multicolumn{1}{c}{}& True Pos. & False Pos. & True Neg. & \multicolumn{1}{c}{ False Neg.} \\
        \cline{2-5}
         CI & 0 & 0 & 8 & 0 \\
         CSE & 9 & 1 & 3 & 0\\
         EFCI & 0 & 0 & 9 & 0\\
         RTI & 1 & 0 & 3 & 0\\
         SAI & 0 & 0 & 13 & 1\\
         URE & 2 & 0 & 0 & 0\\
         \cline{2-5}
    \end{tabular}
\end{subtable}
\end{table}

\subsection{Mutation Methodology}
\label{sec:mutation-methodology}
We first identified scripts that were eligible for this analysis. We modified the \planout{}-A scripts that raised errors when it was appropriate to do so. For example, we updated a number of the scripts that erroneously raised causal sufficiency errors so that they would not raise those errors anymore. We excluded scripts that, for example, contained testing code or configuration code. This allowed us to be reasonably certain that most of the input scripts were correct.

All of our mutations operate over input \planout{} programs, rather than the intermediate representation. We believed this approach would better stress \toolname{}. We perform one mutation per script.

We considered two approaches when deciding how to perform the mutations: 
\begin{enumerate}
    \item\label{method:equal} Randomly select a mutation type, and then randomly select from the eligible AST points for that mutation.
    \item\label{method:weighted} Generate all of the eligible AST points for all of the mutations, and then randomly select from this set. 
\end{enumerate}
Method~\ref{method:equal} leads to an even split between the classes of mutations in the test corpus; method~\ref{method:weighted} leads to frequencies that are proportional to the frequencies of the eligible AST nodes. We chose the latter because we believed it would lead to a more accurate representation of real programming errors. Table~\ref{tab:mut-freq} gives the probability of a script containing a particular mutation type.

To select the subset of scripts to evaluate, we sampled fifty experiments and then selected a random script version from that experiment. We then manually inspected the mutated script and compared the output of the mutation with the original output. 

\paragraph{Findings: Fault Identification over Mutated Scripts}
\label{sec:rq2-findings}
When analyzing our sample of fifty mutated scripts, PlanAlyzer produced only one false positive and only one false negative. The precision and recall were both 92\%. On the one hand, this is very surprising, given both the false positive rate in the \planout{}-A corpus for causal sufficiency errors (8\%) and the proportion of CSE mutations in this sample (28\%). However, we found that most of the CSE mutations caused the program to exit before random assignment, causing \toolname{} to raise legitimate errors about recorded paths with no randomization. The rest were true causal sufficiency errors (i.e., they would cause bias in treatment). 

The one false negative we observed occurred in a script that re-defined the treatment variable for two userids, in what appears to be testing code. The mutation wrapped the redefined treatment, so this is a case where \toolname{} should have raised a ``no randomization error'' in both the input script as well as the mutated script. 

\begin{figure}
    \centering
    \small
    \begin{minipage}[t][6.1cm][t]{0.4\columnwidth}
        \begin{lstlisting}[numbers=left, basicstyle=\incolcode, xleftmargin=2.5em]
...
foo = bernoulliTrial(p=0.4, unit=userid);
if (get_subpop(userid=userid)) {
  if(foo) {
    mode = 2;
    target_level = 3;
    compression_gain = 3;
    lim_enabled = 1;
  }
} else {
  return true;
}
        \end{lstlisting}
    \end{minipage}%
    \begin{minipage}[t][6.1cm][t]{0.6\columnwidth}
        \begin{lstlisting}[numbers=left, basicstyle=\incolcode, xleftmargin=2.5em]
...
randomize = uniformChoice(choices=[0, 1], unit=userid);
if (randomize == 0) {
} else {
    max_delay = uniformChoice(choices=[500, 1000, 2000, 4000, 10000], unit=userid);
    if (max_delay == 500) {
        horizon_seconds = 1;
        max_packets = 9;
    } else if (max_delay == 1000) ...
    } else {
        horizon_seconds = 10;
        max_packets = 166;
    }
    limit_probability = uniformChoice(choices=[1, 5, 10, 20], unit=otherid);
    ...
}
        \end{lstlisting}
    \end{minipage}
    \caption{Example mutations of anonymized scripts from the evaluation set. \emph{Left:} CSE; lines 3, 10-12 have been inserted. This script previously produced valid output, but now raises a causal sufficiency error at the mutation point. \emph{Right:} URE; the unit of randomization in line 14 has been changed from \lstinline{userid} to \lstinline{otherid}. We supply an annotation for \lstinline{otherid} to indicate that it has high cardinality. This mutation raises no errors when producing conditioning sets, but it does raise an error when producing human readable output, saying that the script is not suitable for ATE.}
    \label{fig:mutations}
\end{figure}

\begin{figure}
 \centering
  \hspace*{-1.5em} 
 \includegraphics[width=0.9\columnwidth]{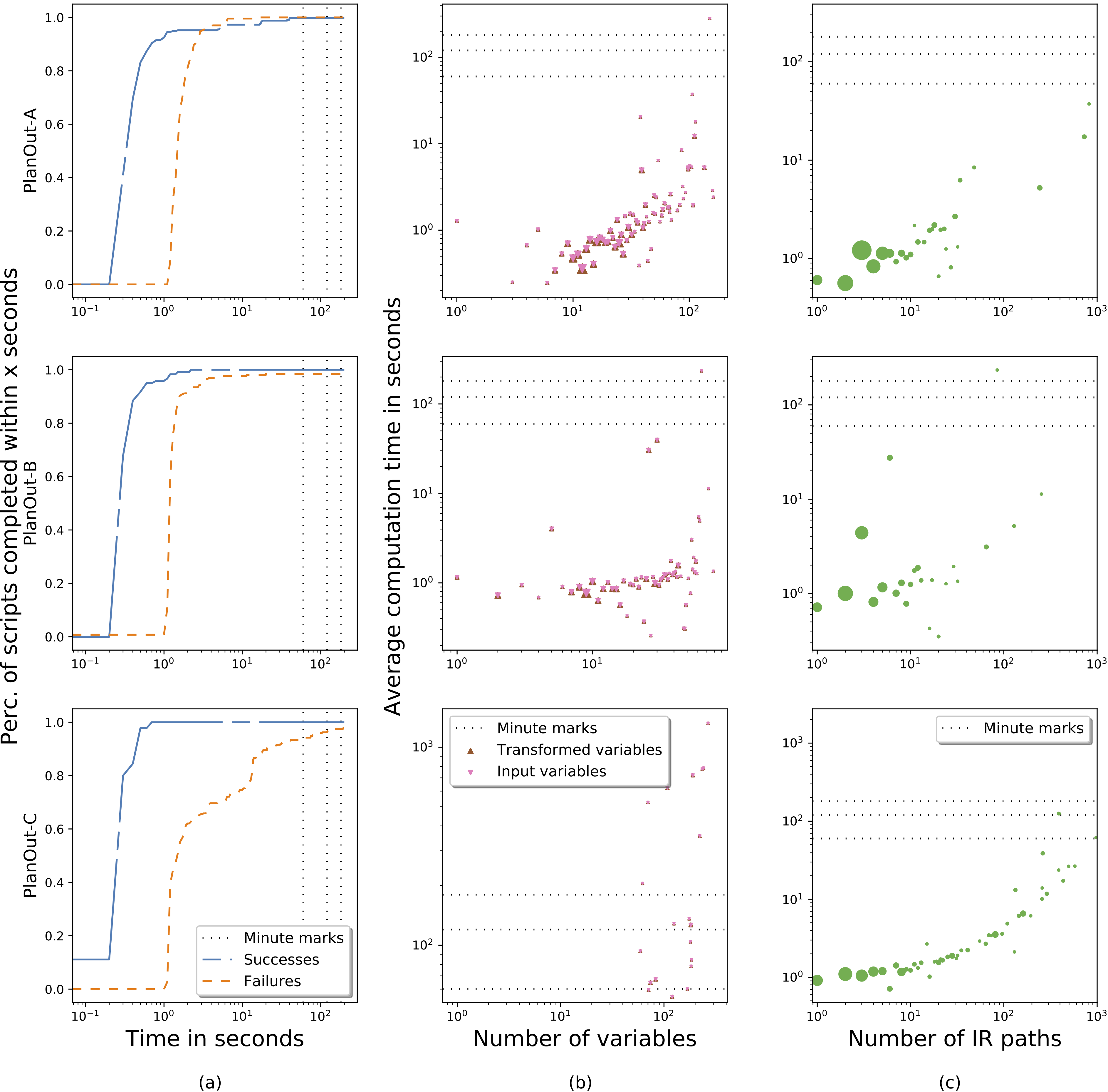}
   \caption{\label{fig:efficiency}Wall-clock timing data for the \planout{} corpus. Plots in column (a) depict the empirical CDF of all scripts on a log-scale. Plots in columns (b) and (c) show the relationship between the running time and features of the \planout{} script we might expect to affect running time, on log-scale on both axes. Plots in column (b) show both the number of variables in the input \planout{} script, and the number of variables in the transformed, intermediate representation of the \planout{} program. Plots in column (c) depict the relationship between the number of paths through \planout{} programs and their running time. The times depicted in both (b) and (c) are averages over scripts satisfying the x-axis value, and the size of the points are proportional to the number of scripts used to compute that average. We chose this representation, rather than reporting error bars, because the data are not iid.}
 \end{figure}


\subsection{Validation against Human-generated Contrasts} 
\label{sec:validation-contrasts}
 We decided whether an experiment should be in the subset according to the following three criteria: (1) all variables in the human-generated contrasts appeared in the original script; (2) \toolname{} was able to produce at least one contrast for the experiment; and (3) \toolname{} produced identical contrasts across all versions of the experiment. Criteria (1) and (2) ensure that analysis does not require knowledge unavailable to \toolname{}. Criteria (3) is necessary because because the tool that monitors contrasts logs them per-experiment, not per-version. If the possible contrasts change between versions, we cannot be sure which version corresponded to the data. Ninety-five of the \numuniqueexps{} unique experiments met these criteria.

\paragraph{Findings: Contrast Generation.} \toolname{} found equivalent contrasts for 78 of the 95 experiments. For 14 experiments, it produced either partial contrasts or no contrasts. In each of these cases, the desired contrast required summing over some of the variables in the program (marginalization), or more sophisticated static analysis than the tool currently supports. Since it is computationally expensive to produce every possible subset of marginalized contrasts, we consider the former to be an acceptable shortcoming of the tool. Finally, 3 experiments had issues with their human-generated contrasts (no contrasts, or ambiguous or unparsable data).
 
\subsection{\toolname{} Performance}
\label{sec:performance}

All analyses were run on a MacBook Air (OSX Version 10.11.6) with a 1.6 GHz Intel Core i5 processor having 4 logical cores. The longest runtime for any analysis was approximately 3 minutes; runtime scales linearly with the number of ``paths'' through the program, where a path is defined according to the transformed internal representation of the input \planout{} program and is related to the number of conditioning sets. \toolname{} uses the Z3 SMT solver~\cite{de2008z3} to ensure that conditioning sets are satisfied and to generate treatments~\cite{valiant1979, fredrikson2014satisfiability}, so both the number of variables in the program and the number of paths in the internal representation could cause a blowup in runtime. We found that runtime increases linearly with the number of internal paths, but possibly exponentially with the number of variables, as depicted in Figure~\ref{fig:efficiency}. 

\paragraph{Findings: Performance} \toolname{} produces meaningful contrasts that are comparable with the human-specified gold standard, automatically generating 82\% of our eligible gold-standard contrasts. \toolname{} runs in a reasonably short amount of time; likely due to \planout{}'s generally small program sizes. 

\subsection{Threats to the Validity of the Evaluation} Although we were able to interview data analysts and experimental designers from several firms that have used \planout{}, we were only able to obtain a corpus from one firm. Although experimentation is ubiquitous and consequential, it can also be contentious, leading firms to be skittish about sharing experimentation scripts. 

One person annotated and verified the scripts, serving as the point of contact with \companyname{} to answer domain-specific questions. The point of contact at \companyname{} is an expert in experimental design. Due to the availability constraints of our contact, we could only verify false positives in the corpus, not false negatives. For the latter, we had to rely on the one annotator reading through the corpus of scripts and using their best judgment. This is a labor-intensive process. As a result, we limited the analyses we did on \planout{}-B and \planout{}-C and resorted to analyzing a sample of the available scripts for mutation testing, which was the most labor-intensive. 

Ideally, our mutations would be defined on the basis of mistakes that people actually make when writing \planout{} scripts. In the absence of that, we defined the mutation operators according to mistakes that we made when learning \planout{} and our experiences working with novices in both programming and experimental design. 

\section{Related Work}
\label{sec:related}
For large firms, concurrent and overlapping experiments complicate the competing goals of enforcing unbiased treatment assignment and preserving high quality experiences for end-users.
Google Layers~\cite{tang2010overlapping} is an experimentation infrastructure that addresses the massive scale of having many different parameters, some of which are not independent and therefore cannot be running concurrently. Kohavi et al. have developed heuristics and best practices for sound experimentation~\cite{crook2009seven, kohavi2009controlled, kohavi2013online, kohavi2015online}.
On a smaller scale, 3X~\cite{ilprints1080} and TurkServer~\cite{parkes2012turkserver} provide open source implementations of complementary resources for experimentation: 3X uses database concepts to store experimental conditions for reproducibility, while TurkServer manages running synchronous experiments on Amazon's Mechanical Turk. 

Systems such as AutoMan~\cite{Barowy:2012:API:2384616.2384663}, SurveyMan~\cite{tosch2014surveyman}, Vox\-PL~\cite{Barowy:2017:VPW:3025453.3026025} and  InterPoll~\cite{livshits2014optimizing} post tasks to crowd-powered backends and address validity questions, or automate statistical results for the user. All use randomization to marginalize some measurement bias, and in some cases estimate power for extremely limited null hypothesis significance tests where the null hypothesis pertains to adversarial respondent behavior. None of these systems are equipped to conduct experiments over a more general hypothesis space.

Two major schools of thought dominate causal inference and influence \toolname{}'s design: the causal graphical models of Pearl~\cite{Pearl:2009:CMR:1642718} and the potential outcomes framework of Neyman and Rubin~\cite{splawa1990application, sekhon2008neyman}. 
Pearl's models have served as inspiration for some probabilistic programming languages~\cite{goodman2008church, anglican}, while many empiricists in economics~\cite{angrist1996identification}, political science~\cite{gelman1990estimating}, and medicine~\cite{little2000causal} have used the potential outcomes framework~\cite{rubin2011causal}. 
Recent work has aimed to unify these two approaches~\cite{hernan2016causalinference, morgan2014counterfactuals}, and to generalize the work across experiments and datasets~\cite{bareinboim2015causal}. 

Prior work on automated analyses of experiments can be found in the information retrieval and knowledge discovery literature. This includes work on the automated search for relationships between data in order to generate testable hypotheses or identify natural or quasi-experiments~\cite{spangler2014automated, sybrandt2017moliere, jensen2008automatic}; tools for specifying causal graphical models and inferring valid queries over them~\cite{scheines2003causal}; and R packages~\cite{tikkaidentifying, gronmping2013doe, gronmping2014frf2, gronmping2014r} and commercial software~\cite{jmp} for identifying causal effects from graphical models, and for generating randomized assignment, replication, and repeated measurements with the appropriate statistical power. None of these approaches or tools treat experiments as software, capable of having errors, or being analyzed.

Spreadsheet analysis occupies a similar problem space as programmatically defined experiments. Both domains feature highly flexible tools (spreadsheet programs and the \planout{} runtime, respectively) where the intent of authors is typically unknown, and where a notion of soundness may not be appropriate. We see parallels between our approach and the development of code smells for spreadsheet features~\cite{hermans2015detecting,hermans2012detecting, cunha2012smellsheet, cunha2012towards} and take inspiration from the qualitative and empirical analyses of spreadsheets~\cite{hermans2015enron, panko1998we}.

\section{Conclusions}
\label{sec:conclusions}
The state of the art for auditing experiments, and for generating their associated statistical analyses, is almost entirely a manual process. 
This is the first work that analyzes field experiments statically. We propose a new class of errors and threats unique to the programmatic specification of experimentation. We have implemented a tool that, for the most common class of experiments, automatically identifies threats and generates statistical analyses. We compare the output of \toolname{} against human-generated analyses of real \planout{} scripts and find that \toolname{} produces comparable results. 

\appendix
\section{Glossary}
\label{app:glossary}
\begin{description}[leftmargin=1em]
    \item[between-subjects] An experiment wherein an experimental subject receives only one treatment. Treatment effect is determined at a population level, by comparing measurements across subjects.
    \item[causal inference] The process of inferring whether one variable can change the outcome of another.
    \item[code smell] A programming pattern that is not an error, but correlates with errors or bad practices, originally defined in Martin Fowler's text on refactoring~\cite{fowler2018refactoring}. Code smells are a popular alternative to fault localization in contexts where faults may be difficult to define, or are tied to programmer intent.
    \item[conditioning set] The set of variables whose values must be fixed to the same set of values for all treatments being compared; can be thought of as a constraint on contrasts.
    \item[contrast] A set of variables that may be compared in order to determine the presence of a causal effect: e.g., treatment and control, A/B/C/etc.
    \item[covariate] A variable that is not a treatment but could be correlated (i.e., could vary) and thus be predictive of outcome. When a covariate is a cause of both outcome and treatment, it is a confounder.
    \item[estimator] A function that estimates the value of a parameter from data. For example, ATE is an estimator for causal effect.
    \item[experimental design] A field of study that focuses on the process of experimentation, from an operational and procedural point of view.
    \item[online field experiment] An experiment conducted over a large, heterogeneous software system, typically involving human interaction, where the treatments are variables in a software system, and the outcome variable of interest is typically a function of human behavior.
    \item[potential outcome] The value of the outcome variable, had treatment assignment been a particular value. Notational convention places the treatment assignment in the superscript of the outcome: $Y^{(\text{treatmentA=a, treatmentB=b, etc.})}$.
    \item[statistical bias] The difference between between the true value of a parameter of interest and the expected value of an estimator of that parameter.
    \item[subgroup analysis] The estimation of treatment effect, split out according to one or more covariates; usually performed when there is heterogeneity in treatment effect.
    \item[treatment] A treatment can refer to any non-empty subset of variable or collection of variables being manipulated, or the value that that variable takes on. In Figure~\ref{code:example-bitrate-planout}, this corresponds to non-empty subsets of  
         \lstinline{dynamic_policy}, \lstinline{max_bitrate} , either variable individually, or the values that either of these may take on, e.g.  \{ \lstinline{dynamic_policy} = false, \lstinline{max_bitrate} = 400 \} or \{ \lstinline{dynamic_policy} = true \}. 
    \item[treatment assignment] The process of \emph{how} a variable (e.g. the label or left-hand side of a variable assignment) is assigned a value. In order to estimate causal effect, treatments must either be assigned in an unbiased manner (e.g., randomly), or correct for biased assignment after data has been collected, during analysis. In the context of programmatically defined experiments, \emph{treatment assignment} refers to the function that maps treatment variables to their values. 
    \item[within-subjects] An experiment wherin the experimental subject receives multiple treatments, and can function as their own control. 
\end{description}

\section{Discussion of \planout{} Features that Complicate Static Analysis}
\label{app:improvements}
We have tagged these with a kind of impact assessment. Each feature may have high (H), medium (M), or low (L) impact on our ability to statically analyze scripts. The projected impact is in terms of correctness, performance, and maintainability of the code. In each case, we explained where and why it causes problems, how it might have an impact on future versions of the tool and extension, and how we can fix it. 

\paragraph{\textbf{Tests (H)}}
It would be helpful to have some kind of directive or pragma at the top of \planout{} scripts to differentiate cases where people are essentially writing configuration files (where they should really be using a different tool) versus test cases, where experiment authors might want to try out versions of the experiment deterministically first.

\paragraph{\textbf{Unit of Analysis (H)}}
We assume that units of randomization are units of analysis. We return the same ATE for a \texttt{userid} as we do for a \texttt{clusterid}. One thing that would help is to explicitly set the unit of analysis at the top of a script, especially in cases where it should be different from the unit of randomization.

\paragraph{\textbf{Types (M/L)}}
There are two core issues with types: (1) there are type errors in some programs, and (2) we need type information to run the SMT solver. 

Regarding (1), in practice most of the type inconsistencies appear in treating one of the external operators as a string versus a container. Types are critical features for end-users. Type errors have often obfuscated more interesting errors in a program. In practice, we have corrected obvious type errors in scripts in order to get to those more interesting errors.  (2) is more critical for the tool, but occurs only in certain cases. Since all built-in random assignment operators must explicitly state their possible outcomes, we only have type issues when we are really badly missing information. In these cases, it is fine for the tool to come back and say it can't generate estimators, but that the script is not invalid. However, when we use data from guards, those often come from external operators having unknown codomains. Types would be very useful for these.

\textit{Potential solution}: add some lightweight type annotations to \planout{} and add them into the serialized \planout{} as metadata.  

\paragraph{\textbf{Default null (M)}}
There is no undefined type in \planout{}, but there is \texttt{null}. This interpretation has been informed by both the Python reference implementation and the execution model as an embedded interpreter in JavaScript. It is not stated anywhere whether the containing environment (e.g., the JavaScript engine in the browser) is expected to have some a particular semantics for when the variable attempts to bind to a variable not defined in the \planout{} script. If a variable is used before definition, we interpret this to mean that the containing environment could either have a concrete value bound to the variable, or the variable could remain null. However, it is also possible that we could call it undefined behavior and throw an error. One of the complexities currently in the static analyzer is that both cases are considered. Note that we always assume that external functions used in the program are defined. 

\textit{Potential Solution}: Formally define and release a semantics for \planout{} as an embedded language.

\paragraph{\textbf{Open-world assumption/Partial evaluation/Concolic evaluation (H)}}
This is related to the previous issue. Since we have an open world assumption, variables that are used but not defined are not errors. \toolname{} runs ``concolic'' (i.e., concrete and symbolic) execution -- when it can enumerate finite values for a variable, it will~\cite{sen2005cute}. When it can't, it will leave the expression symbolic, reducing and normalizing as it goes. 

Since not every variable can be evaluated down to a final value, we need to store these partially-computed values. When we run the SMT solver, we may have insufficient information to concretize the remaining symbolic values. When that happens, we keep the expression symbolic and return that as the ``else'' case alternative for the concrete values.

\textit{Potential Solutions}: The open-world assumption is one of the key features in promoting \planout{}'s adoption. The main challenge in the open-world assumption is just that evaluation can be tricky, there are edge cases, and generating possible values is time-intensive. \toolname{}{} currently only concretizes a variable if there is a unique solution; otherwise, it leaves the expression symbolic. There may be further optimizations and heuristics to the tool that could help with the latter issue. 

\paragraph{\textbf{No knowledge of hypotheses (M)}}
A valid \planout{} script implies a finite set of analyses. Right now we can only produce a subset of those analysis. However, even that subset often contains estimators that are not interesting to the \planout{} user. Systems at e.g., Facebook, already have a mechanism for specifying ``public parameters,'' for more fine-grained logging and monitoring. There is currently not way to couple tracking outcome (i.e., $Y$), so \toolname{} completely abstracts over effects. 

\textit{Potential Solution}: \toolname{} can produce out actual SQL and/or R code if we have some additional information. This feature would be very useful to \planout{} end users.\footnote{e.g., \url{https://github.com/facebook/planout/issues/119}} \planout{} could support e.g., a pragma for specifing the hypothesis we wish to test. This could serve as documentation for the program. It could also connect with some existing formalisms for databases to formally specify experiments~\cite{spivak2014category, spivak2012ologs}. 

\paragraph{\textbf{No user-defined types (L)}}
User-defined types, especially disjoint unions, would be a nice feature to have. Some of the functionality currently in the \toolname{} tool could then be pushed into the SMT solver.

\section*{Acknowledgements}
This material is partially based upon work supported by the United States Air Force under Contract No, FA8750-17-C-0120. We thank Facebook, Inc. for providing the \planout{} corpus and additional funding. Any opinions, findings and conclusions or recommendations expressed in  this material are those of the author(s) and do not necessarily reflect the views of the United States Air Force, nor Facebook, Inc.  Additionally, we thank KDL members Kaleigh Clary, Amanda Gentzel, and Reilly Grant, as well as Sam Baxter, Javier Burroni, John Foley, Cibele Freire, and Bobby Powers for comments, feedback, and suggestions throughout the various incarnations of this work. We thank Dean Eckles for early feedback on this work. Finally, we thank our OOPSLA reviewers, whose detailed comments substantially improved the clarity of this paper.

\bibliographystyle{ACM-Reference-Format}
\bibliography{ref}

\end{document}